\documentclass[a4paper,fleqn]{caswenjia}
\usepackage[numbers]{natbib}
\usepackage{geometry}
\usepackage{amsmath}
\usepackage{amsfonts}
\usepackage{amssymb}
\usepackage{bbm}
\usepackage{hyperref}

\usepackage{caption}

\usepackage{amsmath}
\def\tsc#1{\csdef{#1}{\textsc{\lowercase{#1}}\xspace}}
\tsc{WGM}
\tsc{QE}
\tsc{EP}
\tsc{PMS}
\tsc{BEC}
\tsc{DE}

\begin{document}
\begin{sloppypar}

\let\WriteBookmarks\relax
\def\floatpagepagefraction{1}
\def\textpagefraction{.001}
\shorttitle{Guidance Disentanglement Network for Optics-Guided Thermal UAV Image Super-Resolution}
\shortauthors{Zhicheng Zhao et~al.}

\title [mode = title]{Guidance Disentanglement Network for Optics-Guided Thermal UAV Image Super-Resolution}                      




\author[1,2]{Zhicheng Zhao}
\address[1]{Anhui Provincial Key Laboratory of Multimodal Cognitive Computation, Anhui University, Hefei 230601, China}
\author[1]{Juanjuan Gu}
\author[1,2]{Chenglong Li}
\cormark[1]
\ead{lcl1314@foxmail.com}
\author[1,2]{Chun Wang}
\author[3]{Zhongling Huang}
\author[1]{Jin Tang}
\address[2]{Information Materials and Intelligent Sensing Laboratory of Anhui Province, China}
\address[3]{The BRain and Artificial INtelligence Lab (BRAIN LAB), School of Automation, Northwestern Polytechnical University, Xi’an, 710072, China}
\cortext[cor1]{Corresponding author}
\begin{abstract}
Optics-guided Thermal UAV image Super-Resolution (OTUAV-SR) has attracted significant research interest due to its potential applications in security inspection, agricultural measurement, and object detection. 
Existing methods often employ single guidance model to generate the guidance features from optical images to assist thermal UAV images super-resolution. 
However, single guidance models make it difficult to generate effective guidance features under favorable and adverse conditions in UAV scenarios, thus limiting the performance of OTUAV-SR.
To address this issue, we propose a novel Guidance Disentanglement network (GDNet), which disentangles the optical image representation according to typical UAV scenario attributes to form guidance features under both favorable and adverse conditions, for robust OTUAV-SR.
Moreover, we design an attribute-aware fusion module to combine all attribute-based optical guidance features, which could form a more discriminative representation and fit the attribute-agnostic guidance process. To facilitate OTUAV-SR research in complex UAV scenarios, we introduce VGTSR2.0, a large-scale benchmark dataset containing 3,500 aligned optical-thermal image pairs captured under diverse conditions and scenes. Extensive experiments on VGTSR2.0 demonstrate that GDNet significantly improves OTUAV-SR performance over state-of-the-art methods, especially in the challenging low-light and foggy environments commonly encountered in UAV scenarios.
The dataset and code will be publicly available at \url{https://github.com/Jocelyney/GDNet}.
\end{abstract}

\let\printorcid\relax
\maketitle

\section{Introduction}
{U}{nmanned} Aerial Vehicle (UAV) thermal imaging technology, which can capture infrared radiation emitted by objects, has emerged as a research hotspot in remote sensing due to its unique advantages in various applications, such as security inspection\cite{zhou2022automatic}, agricultural measurement\cite{peng2023accurate}, and object detection\cite{zhou2024automated}. 
The all-weather and all-time operational capability of thermal imaging enables UAV to capture critical information in challenging environments. 
However, constrained by the physical imaging mechanism and technological limitations of onboard thermal imaging devices, the raw thermal images acquired by UAV often suffer from low resolution. 
Consequently, these thermal images lack the rich texture details and fine structures present in optical images, hindering their ability to meet the quality requirements for subsequent high-level semantic understanding tasks. 
This limitation highlights the need for effective thermal UAV image super-resolution techniques to bridge the gap between raw low-resolution thermal images and the desired high-resolution images with enhanced texture details.

To improve the resolution and quality of thermal images, researchers have developed various Single Image Super-Resolution (SISR) algorithms~\cite{zhang2018infrared,liang2021swinir,zamir2022restormer,qiu2023cross}. 
However, existing SISR methods often struggle to accurately reconstruct fine details and textures in thermal images. 
This difficulty arises from the inherent characteristics of thermal images, i.e. the lack of distinct edge information and rich texture. 
Thermal images are smooth and low-contrast, leading to limited high-frequency information. 
As a result, current SISR approaches face significant challenges in recovering intricate details and producing high-fidelity reconstructions, especially in complex UAV environments where image quality is further deteriorated by factors such as atmospheric interference and sensor limitations.

To overcome the limitations of SISR, recent methods~\cite{almasri2018rgb,gupta2020pyramidal,gupta2021toward,zhao2023thermal,zhao2024modality} utilize high-resolution optical images to guide the thermal image super-resolution process.
This Optics-guided Thermal UAV image Super-Resolution (OTUAV-SR) method leverages the superior texture and edge information inherent in optical images, demonstrating significant improvements in reconstructing fine details and enhancing image quality. 
However, although these OTUAV-SR methods can produce excellent super-resolution results, there are still shortcomings. 

\begin{figure}
    \centering
    \includegraphics[width=.92\linewidth]{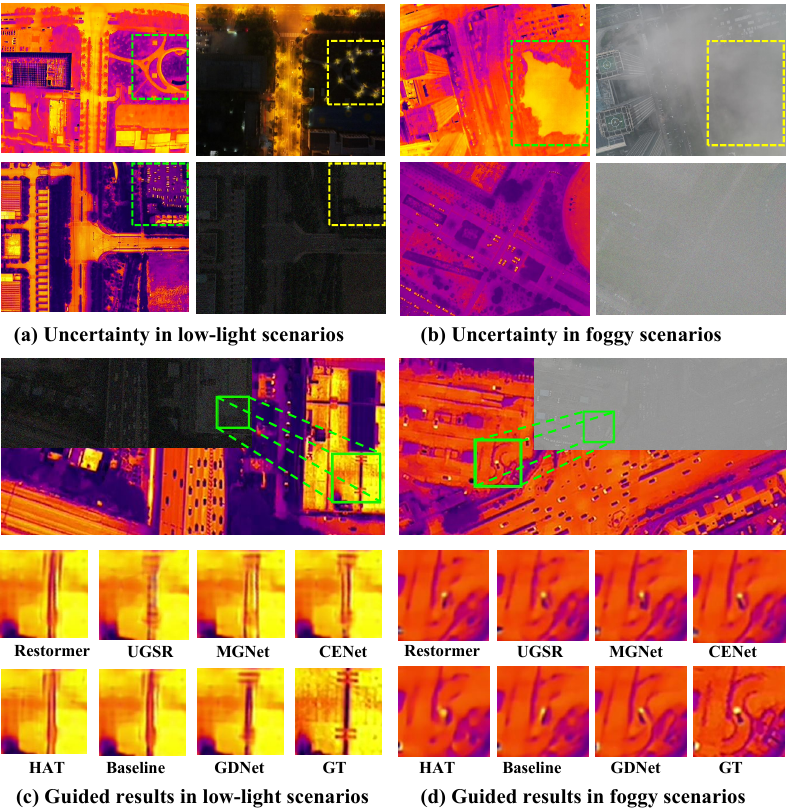}
    \caption{Uncertainty of the optical images in foggy and low-light conditions. We compare the performance 
 of the baseline SwinIR, single image super-resolution methods such as Restormer and HAT, as well as existing guided super-resolution methods, including UGSR, CENet, and MGNet, under low-light and foggy conditions. Our method effectively integrates information from both modalities, leading to superior results.}
    \label{motivation}
\end{figure}

One notable issue is that these methods rely heavily on ideal, high-quality optical image inputs, which are not always available in practice. In challenging UAV scenarios, such as fog or low-light conditions, the performance of existing OTUAV-SR methods degrade significantly.
As illustrated in Fig.~\ref{motivation}(a) and (b), optical images captured by UAV under nighttime and foggy conditions exhibit significant uncertainty,
making it difficult to discern roads and vehicle clusters with low-light and streetlight interference (Fig.~\ref{motivation}(a)), as well as lakes obscured by dense fog (Fig.~\ref{motivation}(b)). 
Utilizing these degraded optical data probably results in inaccurate texture information and artifacts in reconstructed thermal images. 
In some cases, these results are even inferior to those produced by SISR methods. 
As depicted in Fig.~\ref{motivation}(c), the optics-guided thermal super-resolution method fails to accurately recover texture information under low-light condition. In scenarios with dense fog, as shown in Fig.~\ref{motivation}(d), both SISR and guided super-resolution methods cannot reconstruct the thermal image of a curved road. 

Additionally, many OTUAV-SR methods employ a uniform framework across all scenarios, which results in models that cannot respond effectively to different environmental challenges. 
In challenging conditions, existing approaches not only fail to leverage optical guidance information effectively but also easily introduce errors and artifacts.
These findings highlight that current approaches lack the necessary scene-adaptive capability and are not sufficiently robust in complex UAV environments. 
Therefore, developing scenario-specific enhancement techniques in challenging UAV scenarios emerges as a research priority.

In this paper, we propose a novel Guidance Disentanglement Network (GDNet) for robust OTUAV-SR. GDNet disentangles the optical image representation according to typical UAV scenario attributes, then generates effective guidance features under both favorable and adverse conditions. 
Optical images captured by UAV often encounter issues
like low illumination and fog obstruction. To address these issues, we propose an Attribute-specific Guidance Module (AGM), comprising three different guidance branches. We employ a Retinex decomposition network \cite{cai2023retinexformer} to enhance low-light optical images and design an attention block for denoising. Additionally, we introduce a gating attention module to guide the network in focusing on features related to atmospheric haze. These modules leverage the advanced modeling capabilities of the transformer architecture, enabling the network to adapt to varying lighting and visibility conditions.
To adaptively aggregate features from multiple guidance branches,
we introduce the Attribute-aware Fusion Module (AFM). This module enables the network to selectively activate and utilize the most relevant feature branches based on learned representations, ensuring robustness across diverse scenarios. 
Inspired by recent studies \cite{huang2021shuffle,patel2022aggregating,chen2023activating} in window-based self-attention methods, we design an Overlapping Multi-head Cross-Attention Layer (OMCL) to enhance cross-window thermal information connections and expand the receptive field of optical information, facilitating more effective integration of multimodal information. 
To support research in OTUAV-SR under complex UAV scenarios, we introduce VGTSR2.0, a large-scale, well-aligned multimodal UAV dataset, which comprises 3,500 accurately aligned pairs of optical and thermal images captured across diverse scenes, altitudes, lighting conditions, and fog levels. Extensive experiments on VGTSR2.0 demonstrate the superiority of our proposed method.
In summary, the main contributions of this paper are as follows:
\begin{itemize}
\item We propose the Guidance Disentanglement Network (GDNet), a novel framework that decomposes optical image representations into three typical UAV scenarios. GDNet adaptively integrates information across these scenarios to improve thermal image super-resolution (SR) performance. This approach effectively extracts and leverages informative cues specific to challenging conditions, addressing the limitations of existing single-guidance models in diverse UAV environments.

\item We propose the Attribute-specific Guidance Module (AGM), which consists of three distinct structures to model optical representations across various scenarios. Additionally, we introduce the Attribute-aware Fusion Module (AFM) and a training strategy designed to adaptively aggregate features from multiple guidance branches.

\item To facilitate research on OTUAV-SR in complex UAV scenarios, we introduce VGTSR2.0. To the best of our knowledge, VGTSR2.0 is the largest well-aligned multimodal UAV dataset specifically designed for OTUAV-SR tasks. Compared to VGTSR1.0, VGTSR2.0 significantly expands both the dataset's size and diversity to better represent real-world UAV environments. 

\item Extensive quantitative and qualitative experiments on VGTSR2.0 demonstrate that our GDNet significantly outperforms existing state-of-the-art SISR and OTUAV-SR methods. 

\end{itemize}

\section{Related Work}
\label{sec:Related}
In this section, we review the studies most relevant to our research, including single image super-resolution methods and guided image super-resolution methods.

\subsection{Single Image SR Methods}

Single image super-resolution (SISR) aims to reconstruct degraded low-resolution (LR) images into corresponding high-resolution (HR) images without additional auxiliary modalities.
Early approaches are based on prior knowledge of mathematics, such as Random Forest theory~\cite{liu2017image}, regression theory~\cite{he2011single}, and dictionary learning~\cite{yang2012coupled}. 
However, these traditional prior-based methods are limited in their ability to extract and learn features, resulting in a lack of robustness and effectiveness when restoring complex images across diverse scenes.
The introduction of Convolutional Neural Networks (CNNs) significantly advanced SISR technology. SRCNN~\cite{dong2015image} first introduced CNNs to the field of super-resolution reconstruction, achieving better results than traditional methods. While early CNN-based methods~\cite{haris2018deep,zhang2018residual,dai2019second,lim2017enhanced,zhang2018image,niu2020single} primarily focused on optical image SR tasks, they also have potential for thermal image super-resolution.
CNNs use shared weighted convolutional kernels that ensure local connectivity and translation invariance, but they are constrained by the intrinsic locality of convolution operations, limiting their abilities in capturing large-scale structures and modeling long-range dependencies.
To address this limitation, researchers introduced the Transformer~\cite{vaswani2017attention}, which uses self-attention mechanisms to capture global dependencies within images. 
This broader perspective enables the detection of large-scale structures and intricate details, effectively overcoming the limitations of the local receptive fields of CNNs.
Specifically, Liang et al.~\cite{liang2021swinir} proposed SwinIR based on the Swin Transformer~\cite{liu2021swin}, which achieves further improvements in the SISR task.
Chen et al.~\cite{chen2023activating} proposed a novel Hybrid Attention Transformer (HAT) model that combines the channel attention mechanism and self-attention mechanism to activate more input pixels for reconstruction.
TTST~\cite{xiao2024ttst} introduced a selective Transformer for efficient remote sensing image super-resolution.

With the introduction of GAN~\cite{goodfellow2020generative}, researchers found that GAN performs superiorly in generative tasks. Subsequently, GAN-based methods have gradually been applied to image super-resolution tasks. 
For example, Ledig et al.~\cite{ledig2017photo} introduced SRGAN, a pioneering GAN-based method for single image super-resolution.
EEGAN~\cite{jiang2019edge} enhanced GAN-based SR by introducing an edge-enhancement subnetwork, leading to clearer and more robust satellite image reconstruction.
Kong et al.~\cite{kong2023super} developed a dual GAN approach to upscale historic Landsat images for enhanced vegetation monitoring with improved spatial resolution.
Recently, Mamba~\cite{guo2024mambair} and Diffusion models~\cite{rombach2022high} have gained significant attention and have been widely applied in remote sensing. EDiffSR~\cite{xiao2023ediffsr} employed a diffusion probabilistic model for efficient and accurate remote sensing image super-resolution, outperforming other models in both visual quality and computational efficiency.
FMSR~\cite{xiao2024frequency} introduced a state space model tailored for large-scale remote sensing image super-resolution, integrating frequency and spatial cues to enhance reconstruction quality. 

Despite these advancements, reconstructing high-frequency details from low-resolution images remains challenging. This challenge is particularly pronounced in UAV images, where complex operational scenarios further reveal the limitations of existing SISR methods. Developing algorithms capable of reliably recovering fine details and maintaining robust performance across diverse conditions remains imperative.

\begin{figure*}
    \centering
    \includegraphics[width=1\linewidth]{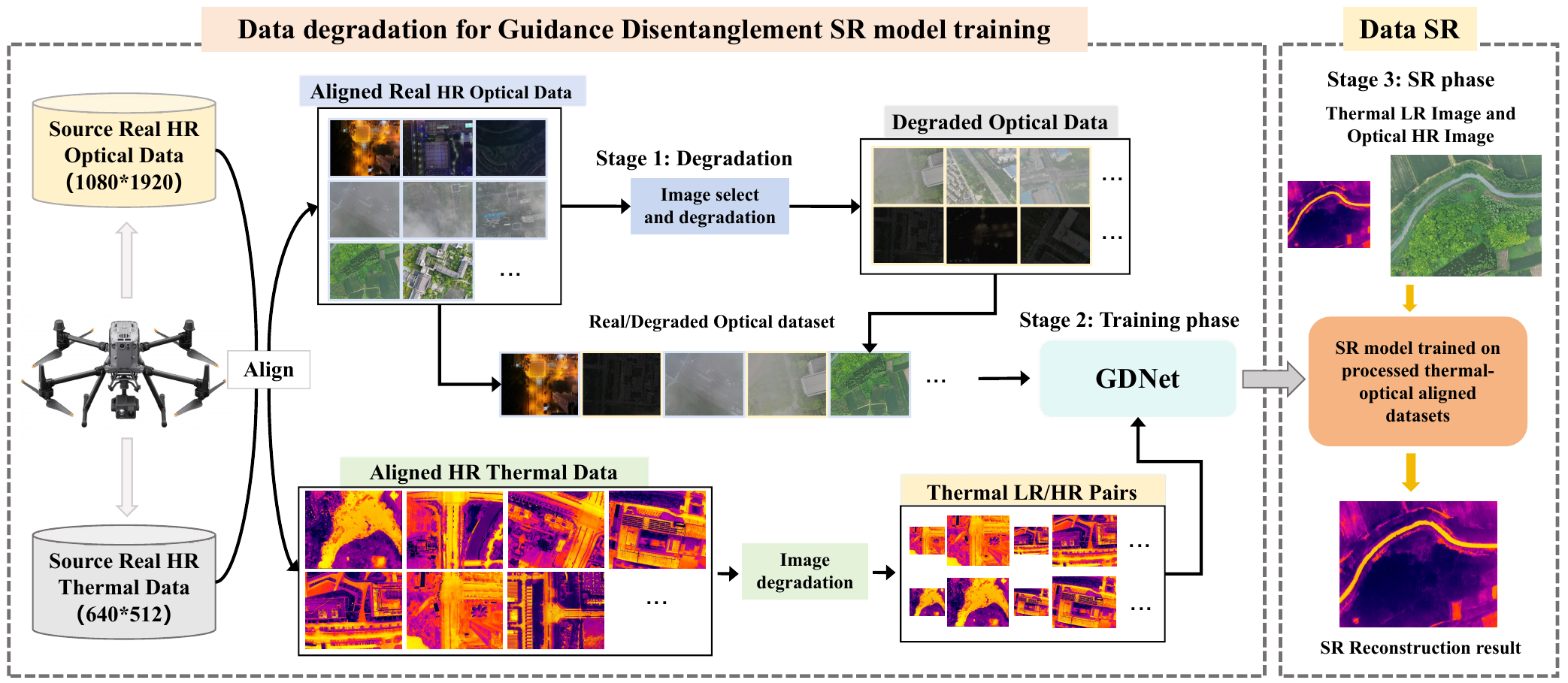}
    \caption{The framework of the proposed guidance disentanglement SR method consists of three stages: Stage 1 involves data acquisition and degradation; Stage 2 focuses on training GDNet using the degraded data; and Stage 3 tests the model by inputting data and evaluating the SR results.}
    \label{framework}
\end{figure*}
\subsection{Guided Image SR Methods}
To address the inherent limitations of single image super-resolution, guided super-resolution methods have been proposed as a solution, which effectively improve image resolution and visual quality by utilizing additional information or prior knowledge.

As the first work to propose guided super-resolution,
Lutio et al.~\cite{lutio2019guided} conceptualized it as a pixel-to-pixel mapping task.
In subsequent developments, super-resolution techniques primarily relied on CNN architectures.
Han et al.~\cite{han2017convolutional} proposed a CNN-based SR algorithm for near-infrared images in low-light environments using optical images.
Fu et al.~\cite{fu2019hyperspectral} proposed a guided SR method that requires no pre-training or supervision.
Chen et al.~\cite{chen2016color} used a color-guided thermal image super-resolution algorithm. 
In addition, other studies~\cite{guo2018hierarchical,wen2018deep,song2020channel} have reconstructed high-resolution depth maps from low-resolution results with the assistance of paired high-resolution RGB images. 
However, these CNN-based methods primarily perform a simple fusion of different modal features, limiting the network's ability to recover image details and structure.
To enhance the model's performance, some research has explored novel architectures that more effectively integrate guided information.
Zhong et al.~\cite{zhong2023deep} proposed an attention kernel learning module that transfers structural information from the guided image to the target image. 
Li et al.~\cite{yanshan2022ogsrn} proposed an optical-guided super-resolution network for SAR images based on the U-Net architecture, which produces results that closely resemble authentic high-resolution SAR images.

In the selection of guidance information, optical images have been favored by many researchers due to rich textural and structural information.
Almasri et al.~\cite{almasri2018rgb} used a GAN-based model to enhance thermal image resolution guided by RGB images. 
Gupta et al.~\cite{gupta2021toward} designed two models to address the alignment of visible images for guiding thermal images. 
Zhao et al.~\cite{zhao2023thermal} employed appearance, edge, and semantic cues in visible images to guide UAV thermal images. 
Additionally, Zhao et al.~\cite{zhao2024modality} proposed a co-learning super-resolution and modality conversion tasks, further enhancing the quality of the generated high-resolution thermal images. 

While current methods have demonstrated the potential to leverage additional information for enhanced image reconstruction, the pursuit of superior image guidance through OTUAV-SR methods remains hindered by the complex and variable conditions encountered in real-world scenes. Rapid changes in lighting, obstruction, and the dynamic nature of aerial imagery necessitate adaptive and robust super-resolution techniques capable of providing consistent guidance under diverse conditions. 

\section{Methodology}
In this section, we first present the framework of Guidance Disentanglement Super-Resolution in Section \ref{overallframework}. Subsequently, Section \ref{arch} provides a comprehensive description of the GDNet model architecture, including the design and implementation details of its key internal modules. Finally, Sections \ref{training} and \ref{loss} discuss the training algorithm and the loss function for GDNet.

\subsection{Overall Framework}
\label{overallframework}
High-quality data is a prerequisite for training effective models. However, according to our survey, there is no UAV thermal super-resolution dataset made specifically for complex scenarios. Therefore, in this work, we first created a new dataset that fulfills the following requirements: (1) \textbf{Positional Consistency.} In UAV scenarios, there are spatial positional differences in the acquired multimodal data due to parallax between cameras. To address this, we first eliminate this positional differences by manual registration; (2) \textbf{Data Balance.} In our dataset, we control the ratio of real scene data to synthetic data to be 2:1, and the ratio of normal light, dark light, and foggy data to be 1:1:1 in all data; (3) \textbf{Data Diversity.} To improve the robustness and generalizability of trained models, a diverse dataset with different degradation levels is crucial. However, acquiring and aligning data with severe degradation poses a significant challenge. 
To overcome this, we select a portion of the well-aligned data and apply synthetic degradation algorithms to obtain severely degraded and aligned data. 

The overall framework, illustrated in Fig.~\ref{framework}, consists of three main stages. After obtaining the raw high resolution (HR) multimodal data captured by the UAV, we manually align the spatial locations across different modal data. First, we perform degradation operations on part of the optical image data and merge them with the real data to produce optical guidance data with multiple degradation levels, which we use as training data. Simultaneously, we conduct downsample operation on high-resolution thermal images to generate training data. Subsequently, we utilize the constructed dataset to train GDNet. Finally, low-resolution (LR) thermal and high-resolution optical images are input into the trained super-resolution model to generate high-resolution thermal images.

\subsection{Data Degradation}
We generate LR-HR thermal image pairs by using bicubic interpolation (BI) and blur-downsampling (BD) degradation models. Users can further degrade high-resolution thermal images with these models to produce low-resolution images that meet specific requirements.

For optical image degradation, low-light images display two primary differences from normal-light images: illumination and noise. To simulate low-light images, we apply a combination of linear and gamma transforms, which can be expressed as follows:
\begin{equation}
I_{LL,i} = \eta_i \times (\zeta_i \times I_{in,i})^{\theta_i}, \quad i \in \{R, G, B\},
\end{equation}

where \(\zeta_i\), \(\eta_i\), and \(\theta_i\) are parameters sampled from uniform distributions: \(\zeta_i \sim U(0.6, 0.9)\), \(\eta_i \sim U(0.3, 0.5)\), and \(\theta_i \sim U(3, 5)\).

Additionally, dark-light scenes often involve noise, which is complex and signal-dependent. The real noise distribution differs significantly from Gaussian distribution. Therefore, we adopt a Gaussian-Poisson noise model \cite{guo2019toward} to simulate both the raw data noise and the noise generated by the photonic sensor. This can be calculated as:
\begin{equation}
I_{out} = \mathcal{R}\left(\mathcal{B}^{-1}\left(\mathcal{P}_{\sigma_p^2}\left(\mathcal{B}\left(\mathcal{R}^{-1}\left(I_{LL}\right)\right) + \mathcal{N}_{\sigma_g^2}\right)\right)\right),
\end{equation}
where $\mathcal{P}_{\sigma_p^2}(\cdot)$ denotes the addition of Poisson noise with variance $\sigma_p^2$, $\mathcal{N}_{\sigma_g^2}$ represents additive white Gaussian noise with variance $\sigma_g^2$, $\mathcal{B}(\cdot)$ is the function converting RGB images to Bayer pattern images, $\mathcal{R}(\cdot)$ is the inverse camera response function.

For haze simulation, we randomly apply algorithms based on atmospheric scattering models and masks to each image to mitigate the overfitting of neural networks. The haze image simulated using the atmospheric scattering principle can be represented as:
\begin{equation}
I_{hazy} = I_{img} \cdot e^{-\beta \cdot d} + A \cdot (1 - e^{-\beta \cdot d}),
\end{equation}
where \( \beta \) is a positive parameter representing the scattering coefficient of the atmospheric medium, \( d \) denotes the distance of each pixel in the image from the center of the haze, and \( A \) represents the intensity of the atmospheric illumination. \( e^{-\beta \cdot d} \) describes the exponential decay with respect to distance.

\begin{figure*}
    \centering
    \includegraphics[width=1\linewidth]{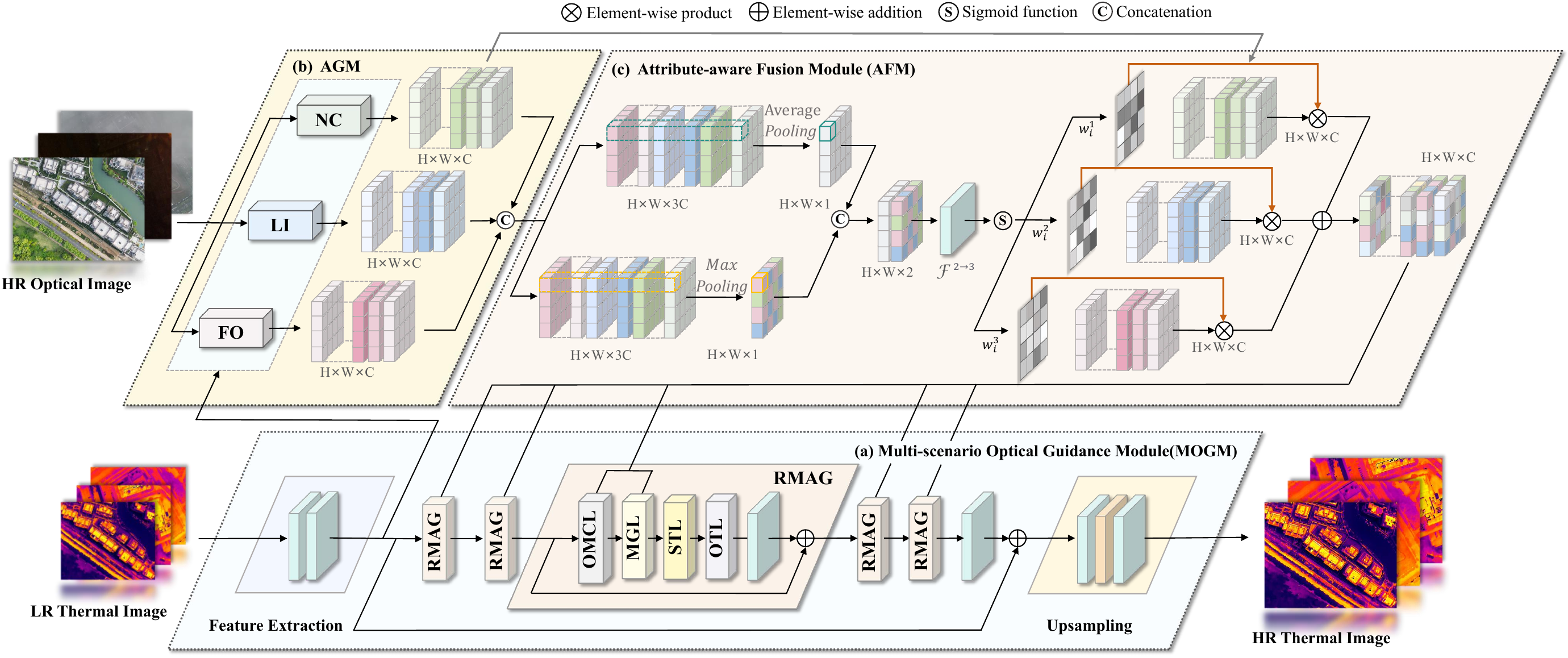}
    \caption{The overall structure of the proposed GDNet. The Attribute-specific Guidance Module (AGM) consists of three branches: NC, LI, and FO, which represent attribute-specific branches for normal conditions, low illumination, and fog obstruction, respectively. To aggregate features from these branches, we introduce the Attribute-aware Fusion Module (AFM), which facilitates the adaptive aggregation of the outputs. Additionally, the MOGM enhances feature representation by integrating both optical and thermal information.}
    \label{overall}
\end{figure*}

\subsection{GDNet Architecture}
\label{arch}
The architecture of the proposed GDNet is shown in Fig.~\ref{overall}. 
The degraded LR-HR thermal image pairs, simulated from the source sensor images and well-aligned real-artificial optical images, are input into GDNet for training.
For thermal images, the initial shallow feature maps are obtained by a feature extractor that consists of a single 3$\times$3 convolutional layer, which also increases the number of channels:
\begin{equation}
F_{initial}=\text{Conv}(X_{LR}) \in \mathbb{R}^{H \times W \times C},
\end{equation}
where \(\text{Conv}(\cdot)\) represents the \( 3 \times 3 \) convolution operation, and \( F_{\text{initial}} \in \mathbb{R}^{H \times W \times C} \) denotes the extracted feature map.
Optical images are simultaneously fed into the pre-trained Attribute-specific Guidance Module (AGM) to obtain three attribute-specific and decoupled representations:
\begin{align}
F_{n} &= f_{NC}(Y) \in \mathbb{R}^{H \times W \times C}, \\
F_{l} &= f_{LI}(Y) \in \mathbb{R}^{H \times W \times C}, \\
F_{f} &= f_{FO}(Y) \in \mathbb{R}^{H \times W \times C},
\end{align}
where \( f_{NC}(\cdot) \), \( f_{LI}(\cdot) \), and \( f_{FO}(\cdot) \) represents the Normal Condition (NC), Low Illumination (LI), and Fog Obstruction (FO) operations in Fig.~\ref{overall}(b), respectively. \( F_{{n,l,f}} \in \mathbb{R}^{H \times W \times C} \) represent the enhanced optical feature maps, which are adept at processing specific attributes.
However, attribute annotations are available during training but absent during testing, creating uncertainty regarding which fusion branch to activate in OTUAV-SR. To address this challenge, we introduce the Attribute-aware Fusion Module (AFM) to effectively aggregate attribute-specific features, as illustrated in Fig.~\ref{overall}(c). 
To establish long-range dependencies effectively and fully utilize the shallow information from thermal images and the enhanced optical guidance information, the Residual Multiple Attention Group (RMAG) is incorporated into the GDNet.
The initial thermal features and the aggregated optical features after AFM are fed into each RMAG, where the feature map of the \(i\)-th RMAG is described:
\begin{align}
F_{i} &= f_{RMAG_i}(F_{i-1}, f_{AFM}(F_{optic})) + W_i F_{initial},
\end{align}
where \( F_{\text{optic}} \) denotes the optical feature map, obtained by concatenating \( F_n \), \( F_l \), and \( F_f \). \( f_{AFM}(\cdot) \) represents the AFM, as shown in Fig.~\ref{overall}(c), \( F_{i-1} \) and \( F_i \) represent the input and output feature of the \(i\)-th RMAG, respectively, \( f_{\text{RMAG}_i}(\cdot) \) denotes the \(i\)-th RMAG, and \( W_i \) is a learnable parameter.
Finally, the output is upsampled to generate the final SR result as follows:
\begin{equation}
X_{SR} = \text{Conv}(f_{pixel-shuffle}\left(\text{Conv}\left(F_{I}\right)\right)),
\end{equation}
where \( F_{I} \) is the output feature of the last RMAG, \(\text{Conv}(\cdot)\) represents the \( 3 \times 3 \) convolution operation, and \(f_{pixel-shuffle}(\cdot)\) denotes PixelShuffle upsampling method.

\subsubsection{Attribute-specific Guidance Module}
\label{agmm}
To enhance the extraction of optical features under varying weather conditions, we propose an Attribute-specific Guidance Module (AGM) comprising three branches with distinct architectures, each designed to address specific challenges. The NC branch facilitates the interaction between optical and thermal features, enabling the extraction of more valuable information. The LI branch aims to improve the visibility of dark areas in low-light conditions based on Retinex theory with transformer architecture. The FO branch selectively preserves detailed information in foggy images, aiding in the restoration of clarity and detail in haze-obscured images.

\noindent\textbf{Normal Condition Branch.}
\label{mgldiscription}
Noise in thermal and optical images captured by UAV under normal lighting condition increases with altitude. To mitigate this issue, we employ a cross-attention mechanism that integrates thermal features. This mechanism facilitates comprehensive interaction between optical and thermal features, thereby reducing atmospheric noise in optical images. The NC branch consists of a basic CNN backbone and two Multimodal Guidance Layers (MGL), which incorporate cross-attention and shift window mechanisms to enhance guidance information. 
\begin{figure}
    \centering
    \includegraphics[width=.98\linewidth]{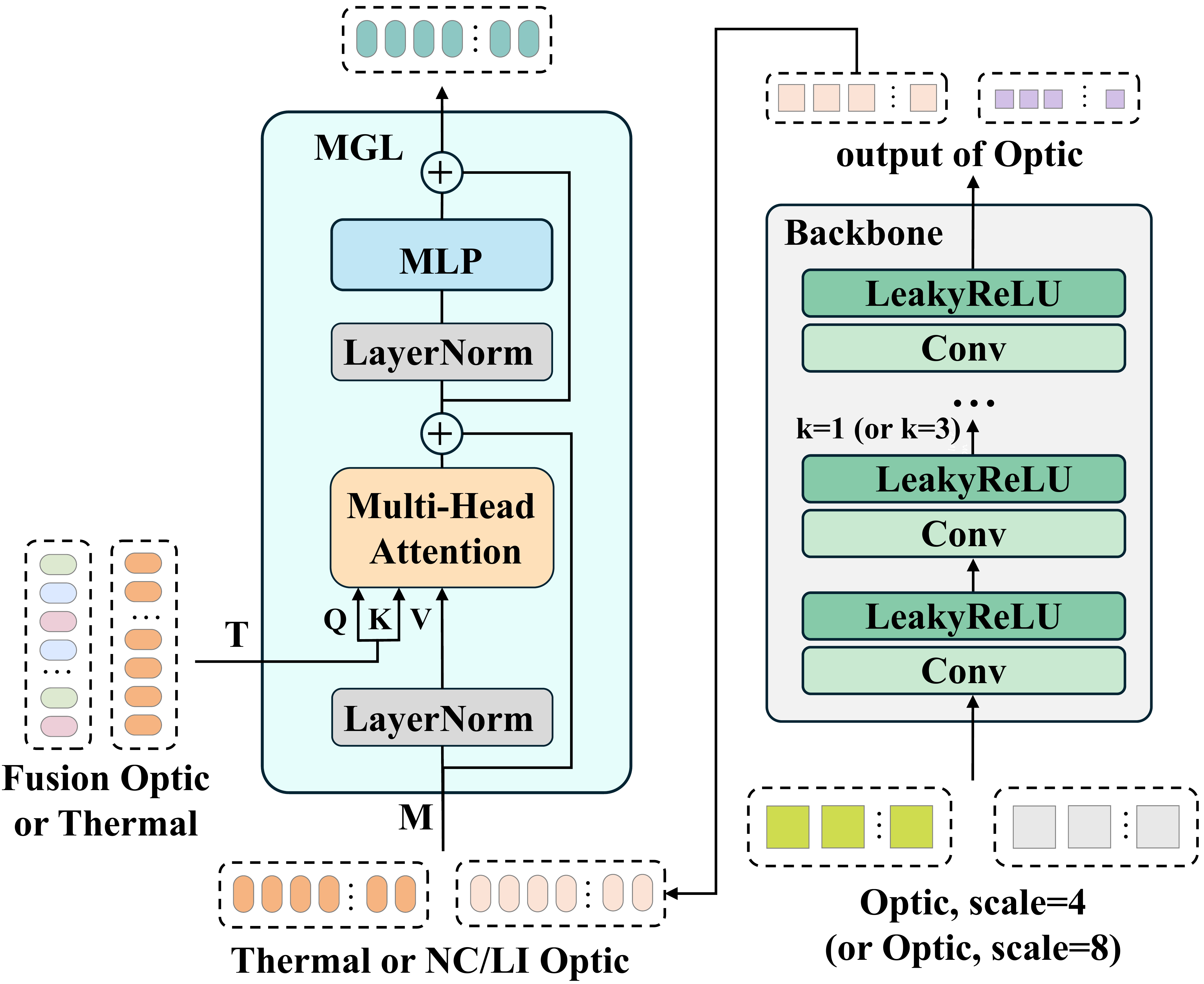}
    \caption{The network structures of Backbone and MGL. Optical images are input into NC and FO, where they first perform shallow feature extraction and size reduction through Backbone, followed by the MGL module.}
    \label{mgl}
\end{figure}
Optical images are initially processed by a basic backbone consisting of multiple 3$\times$3 convolutional layers, followed by LeakyReLU activation functions. Subsequently, the output features are passed through a Multi-head Cross-Attention (MCA) layer employing a sliding window mechanism, as illustrated in Fig.~\ref{mgl}. In MCA, these operations are computed as follows:
\begin{equation}
Q = YW^Q, \; K = YW^K, \; V = XW^V,
\end{equation}
where $W^Q$, $W^K$, and $W^V$ are learnable projection matrices shared across different windows, $X$ denotes the optical features, and $Y$ represents the thermal features. The attention matrix is computed from the self-attention or cross-attention mechanism within the local window as follows:
\begin{equation}
Attention(Q, K, V) = \text{softmax}(QK^T / \sqrt{d_k} + B) V,
\end{equation}
where $B$ is the learnable relative positional encoding matrix, and $d_k$ is the dimension of the query $K$. The $Attention(\cdot)$ denotes cross-attention operation. In MCA, $Q$ and $K$ are derived from the thermal features, while $V$ is derived from the optical features. The complete flow of MGL is as follows:
\begin{flalign}
\begin{split}
 &M       =  {\rm   ({LeakyReLU}({Conv}}(M)))_{\times k}, \\
  &M_{l}^{\prime}=   {\rm MCA }(  {\rm LN } ( M _{l} , T) + M_{l},   \\
  &M_{l+1} =  {\rm MLP }(  {\rm LN } (M_{l} ^{\prime}  )+ M_{l}^{\prime}+ M_{l}, \\
\end{split}
\end{flalign}
where \(k\) represents the number of Conv+LeakyReLU operations, MLP and LN refer to the multi-layer perceptron and layer normalization, respectively, and $T$ denotes thermal features. $M_{l}$ and $M_{l+1}$ denote the input and output features of the MCA and MLP for layer $l$, respectively.

\noindent\textbf{Low Illumination Branch.}
Our LI branch consists of three components: a pre-trained Retinex decomposition network, a basic CNN backbone, and six MGLs instead of using a simple feature extractor. 
The Retinex decomposition network is a flexible network that can be based on any image decomposition model~\cite{cai2023retinexformer,fan2023rme,yi2023diff}. We can train the decomposition network from scratch using paired inputs or simply load publicly available pre-trained weights. After constructing the Retinex decomposition network by directly loading RetinexNet's publicly available pre-trained weights, we observe that the results performed well on synthetic data but underperformed on real data. 
To better simulate the visual scene of a UAV under low-light conditions, we train the decomposition network from scratch based on RetinexFormer~\cite{cai2023retinexformer}, which offers a more robust feature representation by modeling noise in reflectance and illumination maps through a transformer structure. Specifically, we carefully selected one-third of the optical images from the VGTSR2.0 training set and 1000 images from the DroneVehicle~\cite{sun2022drone} dataset. Then, these images are synthesized into pairs under low-light and normal-light conditions using the non-deep Dark ISP method. Additionally, to better approximate the low-light environments encountered in real-world scenarios, we introduced the LOLv2~\cite{yang2021sparse} dataset to enhance the generalization and robustness of the decomposition model.
\begin{figure}
    \centering
    \includegraphics[width=.94\linewidth]{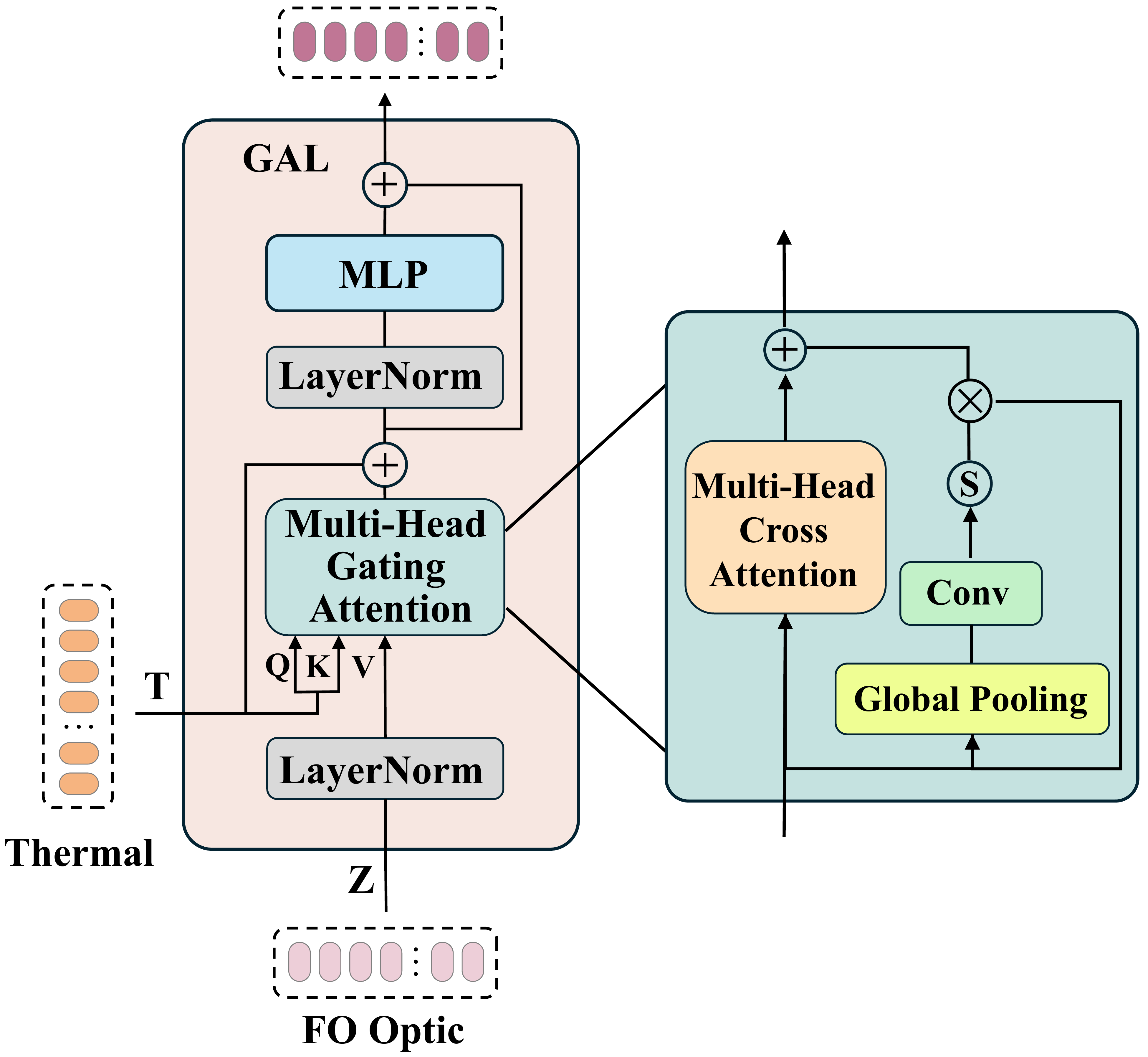}
    \caption{The network structure of GAL. We use a gating mechanism to filter fog-related features from foggy optical images and interact with fog-resistant thermal images to generate more discriminative optical features.}
    \label{gal}
\end{figure}

\noindent\textbf{Fog Obstruction Branch.}
Under foggy condition, the local details of the image become blurred and obscured. To enhance clear representations region and suppress foggy ones, we integrate a gating mechanism into the original Transformer architecture. Our FO branch comprises a backbone and four Gating Attention Layers (GAL), as illustrated in Fig.~\ref{gal}. The spatial dimensions of the input feature map are first compressed into a channel descriptor. A weight matrix is then generated for each channel, reflecting its importance. Finally, the feature maps for each channel are weighted to enhance the critical features. The entire process in FO is formulated as follows:
\begin{flalign}
\begin{split}
 &Z       =  {\rm   ({LeakyReLU}({Conv}}(Z)))_{\times k}, \\
 & Z_{LN}^{\prime} = \text{LN}(Z), \\
&Z_{GAB} ^{\prime}= f_{sig}(Conv(f_{gp}(Z_{LN}^{\prime})))\times Z_{LN}^{\prime},\\
  &Z_{l}^{\prime}=   {\rm MCA }(  {\rm LN } ( Z _{LN} ^{\prime} , T) +Z_{GAB} ^{\prime}+ T, \\
  &Z_{l+1} =  {\rm MLP }(  {\rm LN } (Z_{l} ^{\prime}  )+ Z_{l}^{\prime}+ Z_{LN} ^{\prime}, \\
\end{split}
\end{flalign}
where \( Z_{LN}^{\prime} \), \( Z_{GAB}^{\prime} \), and \( Z_{l}^{\prime} \) represent the intermediate features, the functions \( f_{sig}(\cdot) \) and \( f_{gp}(\cdot) \) denote the Sigmoid and Global Pooling functions, respectively. The MLP and LN refer to the multi-layer perceptron and layer normalization functions, respectively. \( Z_{l+1} \) represents the output features of the MCA module, GAB module, and MLP module for layer \( l \).

\subsubsection{Attribute-aware Fusion Module}
After obtaining three features corresponding to different attributes through the AGM, we employ the Attribute-aware Fusion Module (AFM) to perform adaptive fusion operations on them. Different from SKNet\cite{li2019selective}, which integrates information along channel dimensions, our approach adaptively aggregates these three features in the spatial dimension, as shown in Fig.~\ref{overall}(c). This approach better preserves the spatial structure of feature maps, enabling the model to focus more on the local details of the image while performing adaptive aggregation. The whole AFM process can be described as follows:
\begin{flalign}
\begin{split}
&F_{optic}=Concate(F_{n}, F_{l}, F_{f}),\\
&Attn= Concate(\mathcal{P}_{avg}(F_{optic}),\mathcal{P}_{max}(F_{optic})), \\
&\widehat{Attn_{i}}= f_{sig}(\mathcal{F}^{2\rightarrow 3}(Attn_{i})), \quad i=1,2,3,\\
&S = \mathcal{F}(\widehat{Attn_{1}} \cdot F_{n} +  \widehat{Attn_{2}} \cdot F_{l} + \widehat{Atten}_{3} \cdot F_{f}), \\
\end{split}
\end{flalign}
where \( F_{optic} \) represents the optical feature map, obtained by concatenating $F_{n}$, $F_{l}$, and $F_{f}$, $\mathcal{P}_{avg}(\cdot)$ and $\mathcal{P}_{max}(\cdot)$ denote average and maximum pooling, respectively, and $\mathcal{F}(\cdot)$ represents the convolution applied to the weighted results.

\subsubsection{Multi-scenario Optical Guidance Module}
To facilitate effective interaction between optical and thermal features, we design a  Multi-scenario Optical Guidance Module (MOGM). The MOGM comprises four Residual Multiple Attention Groups (RMAGs), each containing OMCL, MGL, STL, and OTL, as illustrated in Fig.~\ref{overall}(a).
\begin{figure}
    \centering
    \includegraphics[width=0.96\linewidth]{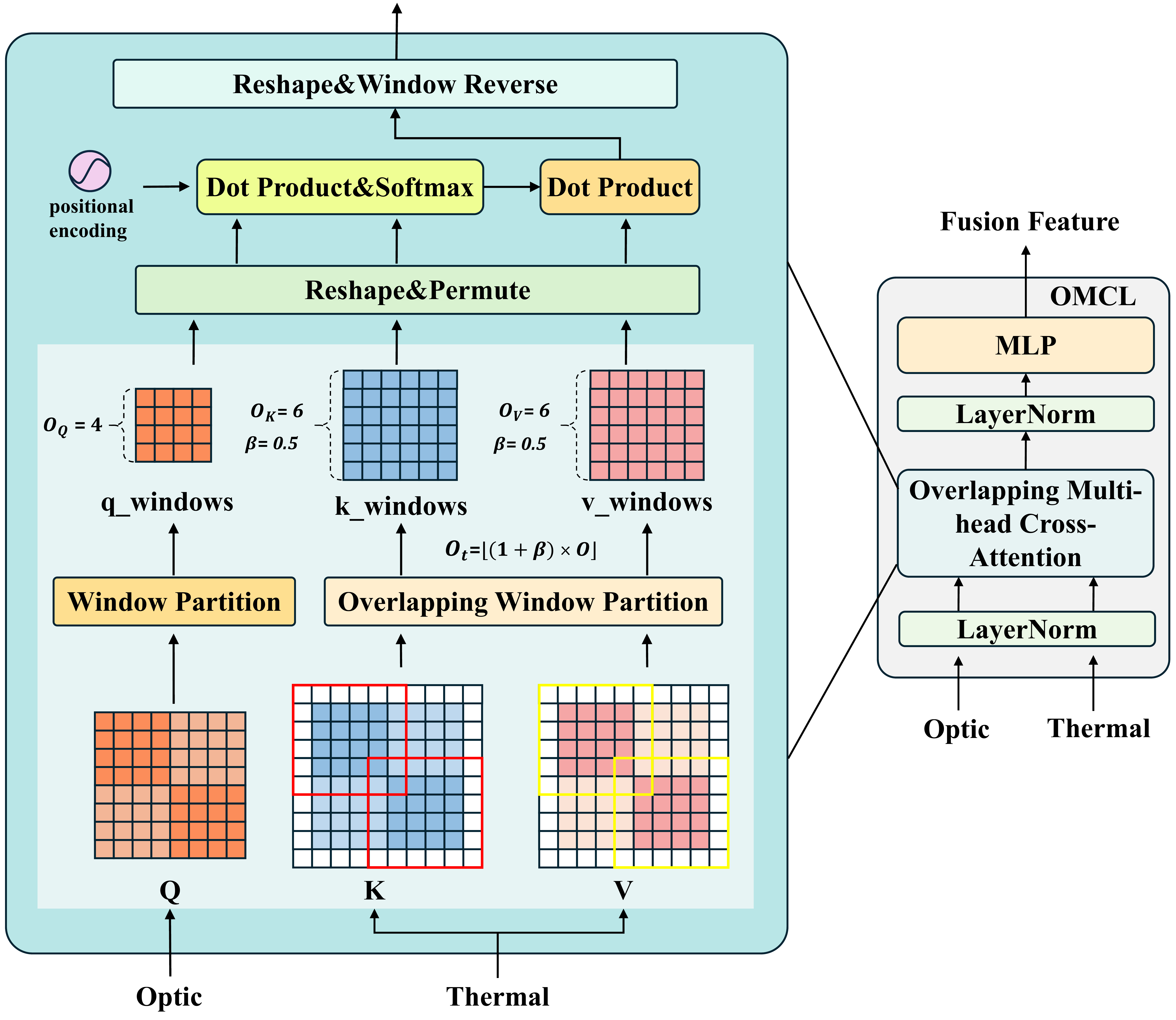}
    \caption{The network structure of OMCL. The q-window from optical features, obtained after ordinary window partition, interacts with the k-window and v-window from thermal images after overlapping window partition, \( O\) denotes the window size, and \( \beta\) represents the window overlapping ratio.}
    \label{omcl}
\end{figure}

\noindent\textbf{Overlapping Multi-head Cross-attention Layer.}  Inspired by the excellent work HAT \cite{chen2023activating}, we introduce the Overlapping Multi-head Cross-attention Layer (OMCL) into the model. In OMCL, we incorporate a multimodal overlapping multi-head cross-attention (OMCA), which ensures optical information can query a larger window and extend window connectivity during the multimodal fusion process, akin to the architecture of the conventional Swin Transformer, as illustrated in Fig.~\ref{omcl}.

\noindent\textbf{Overlapping Transformer Layer.}
We employ the original Overlapping Transformer Layer (OTL) from HAT \cite{chen2023activating}, where the $Q$, $K$, and $V$ in this attention mechanism are derived from the features of a single input modality. Additionally, our ablation experiments demonstrate that OMCL achieves superior performance.

\noindent\textbf{Multimodal Guidance Layer.}
Section \ref{mgldiscription} details the structure and calculation of MGL. The primary distinction lies in the MCA calculation, where $X$ represents thermal features and $Y$ denotes optically fused features. Specifically, $Q$ and $K$ are derived from optical features, while $V$ originates from thermal features.

\noindent\textbf{Swin Transformer Layer.}
To capture long-range dependencies and model global contextual information, the Swin Transformer Layer (STL) employs self-attention and shifted window mechanisms to further enhance the fused information. The entire process in STL is formulated as follows:
\begin{flalign}
\begin{split}
&\hat{t}^l= {\rm  \text{MSA}}({\rm LN}(t^{l-1})) + t^{l-1},\\
&t^l = {\rm MLP}({\rm LN(\hat{t}^l)}) + \hat{t}^l,\\
\end{split}
\end{flalign}
where $t^{l}$ denotes the output of $l$-th STL. MSA denotes the function of multi-head self-attention,
MLP and LN represent the functions of the multi-layer perceptron and layer normalization, respectively.

\begin{figure*}
    \centering
    \includegraphics[width=.95\linewidth]{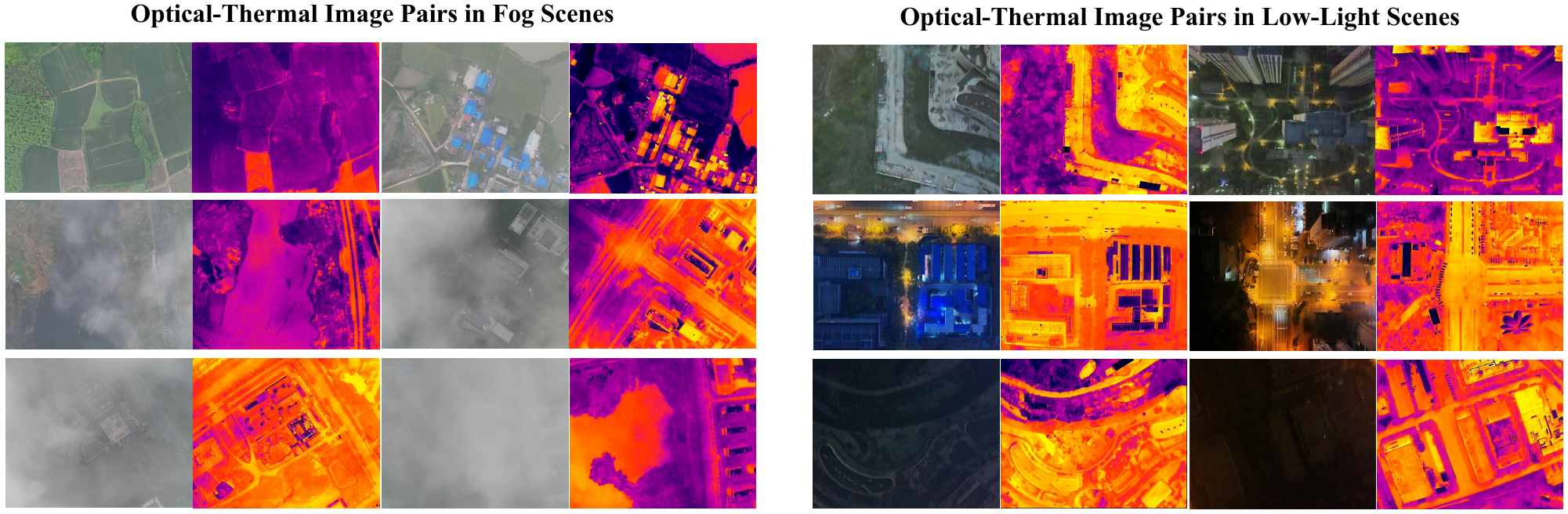}
    \caption{Optical and thermal UAV image pairs in VGTSR2.0. The left side shows pairs of drone optical and thermal images taken under clouds with increasing fog levels. In contrast, the right side presents pairs of optical and thermal images obtained from UAV in low-light conditions, ranging from evening to late night.}
    \label{condition}
\end{figure*}
\begin{figure*}
    \centering
    \includegraphics[width=.98\linewidth]{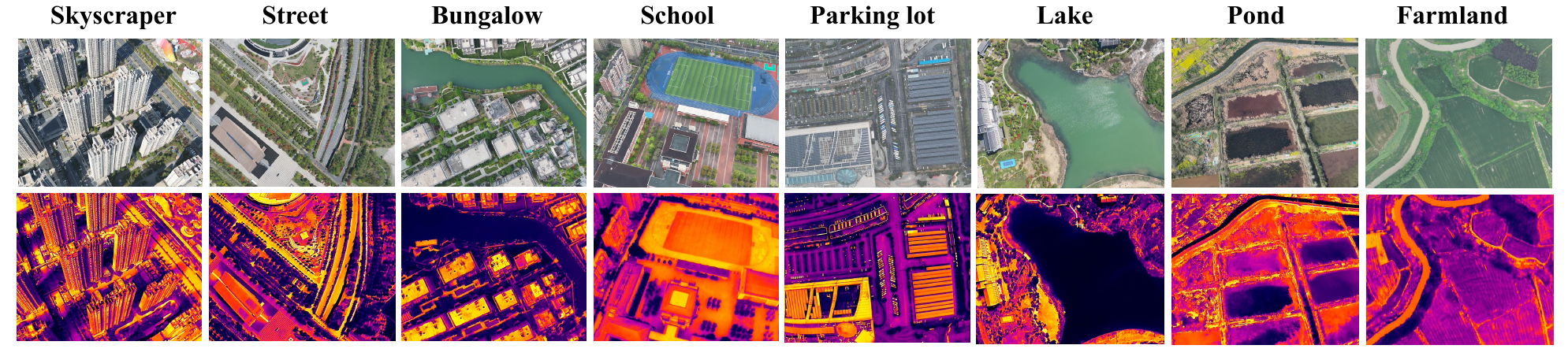}
    \caption{The key scenes in the VGTSR2.0 dataset include Bungalow, Skyscraper, Parking Lot, Farmland, Pond, Lake, Street, and School, all of which were captured at altitudes of at least 200 meters.}
    \label{sence}
\end{figure*}
\subsection{Training Algorithm}
\label{training}
During the training phase, three key issues must be addressed. First, if directly using the entire training set to train the network, the errors introduced by low-light and fog conditions will propagate through all branches. Second, since the scenes are unknowable during the testing phase, predicting the specific challenges posed by optical representations in the OTUAV-SR task is not feasible. Our ultimate goal is to enhance the ability to leverage emerging optical representations while suppressing responses to unrepresented features. To address these issues, we propose a three-stage training strategy to facilitate efficient learning for the guided thermal super-resolution task in complex scenes.

\noindent\textbf{Training of Backbone:}
In the first stage, we use all VGTSR2.0 training data to train feature extraction (as shown as the Backbone in Fig.~\ref{mgl}) for optical images under generic weather conditions. This module includes 4 sets of convolutional layers and LeakyReLU activation functions for $\times$4 SR ,while 6 sets for $\times$8 SR.Note that all three guidance branches and the attribute-aware fusion layer are removed, with the main objective of training a generic feature extractor for all attribute branches.

\noindent\textbf{Training of All Attribute-specific Branches:}
After obtaining the trained feature extractor, we sequentially train the three remaining attribute-specific modules in each branch, using the corresponding attribute-specific training data to learn attribute-specific optical representations. All other parameters are frozen in this process. Before training the LI branch, we pre-train the Retinex Decomposition Network based on RetinexFormer as outlined in Section \ref{agmm}. Note that only data from the corresponding scene are utilized when training a specific branch. Data from other scenes, along with their branch structures, are excluded from the network. The training setup remains consistent with the first stage.

\noindent\textbf{Training of Attribute-aware Fusion Module and Fine-tuning of MOGM:}
After obtaining the learned feature extractor and the three attribute-specific branches, we use all data to train the attribute-aware fusion module and fine-tune MOGM while freezing all other parameters. The learning rates for the attribute-aware fusion module and MOGM are set to 10$^{-4}$ and 0.5$\times$10$^{-4}$, respectively. Other training settings remain unchanged.
\setlength{\floatsep}{5pt plus 2pt minus 2pt}
\setlength{\textfloatsep}{5pt plus 2pt minus 2pt}
\setlength{\intextsep}{5pt plus 2pt minus 2pt}
\subsection{Loss Function}
\label{loss}
The entire network is optimized by minimizing the L1 loss between the predicted UAV thermal image and the real GT thermal image. The loss is calculated as follows:
\begin{equation}
\mathcal{L}(\Phi) = \frac{1}{M} \sum_{j=1}^{M} || F_{out}^{j} - F_{GT}^{j} ||_{1},
\end{equation}
where $\mathcal{L}(\Phi)$ represents the loss function, with $\Phi$ denoting the set of parameters to be optimized within GDNet. $F_{out}^{j}$ refers to the \( j \)-th pixel of the output thermal image produced by GDNet, and $F_{GT}^{j}$ corresponds to the \( j \)-th pixel of the ground-truth thermal image.

\begin{table*}[tb]
\renewcommand{\arraystretch}{1.0}
\begin{center}
\footnotesize
\caption{Detailed comparison of different SR datasets. "Thr" represents thermal images, while "Opt" denotes optical images.}
\begin{tabular}{ccccccccccc}
\toprule
Dataset& Resolution&Sence&Modal &Number&Thr:Real/Synthetic&Opt:Real/Synthetic&Aligned \\\hline
Set14&-&Outdoor+Indoor &Optical &14 &Synthetic &- &-   \\
T91&-&Outdoor &Optical &91 &Synthetic  &- &-  \\
Urban100&-&Outdoor+Indoor &Optical &100 &Synthetic  &- &-  \\
RealSR&-&Outdoor+Indoor &Optical &595 &Real &- &-  \\
Flickr2K&2048×1080&Outdoor+Indoor &Optical & 2650&Synthetic  &- &-  \\
DIV2K&2048×1080&Outdoor+Indoor &Optical & 1000 &Synthetic  &- &-  \\
CATS &1280x960 &Outdoor+Indoor &Optical+Thermal &2744 &Synthetic&Real&$\times$ \\
VGTSR&640×512&Remote Sensing&Optical+Thermal&2050&Synthetic&Real&$\checkmark$ \\
\textbf{VGTSR2.0(ours)}&640×512&Remote Sensing&Optical+Thermal&7000&Synthetic&Real+Synthetic&$\checkmark$ \\ \toprule
\end{tabular}
\label{datacompare}
\end{center}
\end{table*}

\begin{table}[tb]
\renewcommand{\arraystretch}{1.2}
\begin{center}
\caption{The number of training and test sets when training GDNet on VGTSR 2.0, along with a comparison to three scenarios in VGTSR 1.0 (Pairs).
}
\scriptsize
\begin{tabular}{ccc|cccc|cccc}
\toprule
&&VGTSR2.0&&Train&Test&Sum &VGTSR1.0&&  \\ \hline
&&All&&2798&702&3500&1025&&  \\
&&Norm&&926&230&1156&880&& \\
&&Fog&&947&235&1182 &70&&\\
&&Lowlight&&925&237&1162&75&& \\ \toprule
\end{tabular}
\label{train}
\end{center}
\end{table}
\section{DATASET}
In this section, we will introduce our Visible image Guided Thermal image Super-Resolution-2.0 (VGTSR2.0) dataset, including existing datasets, data collection and analysis, as well as dataset challenges.
\subsection{Existing Datasets}
Existing studies on OTUAV-SR often rely on datasets like FLIR-ADAS\cite{gupta2021toward} and CATS\cite{treible2017cats}, which are derived from handheld devices. These datasets provide high-resolution optical and lower-resolution thermal images but suffer from alignment issues. Additionally, they are constrained by the shooting angles and conditions of the devices, making them less suitable for complex scenarios.
The VGTSR1.0 dataset has advanced multimodal image SR, particularly regarding image alignment and quality. However, it is limited in scene diversity and weather conditions, primarily focusing on campus and street scenes, and lacks a substantial number of image pairs under various lighting and fog conditions. To address these limitations and enhance the robustness evaluation of thermal image SR algorithms across diverse environments, we introduce the VGTSR2.0 dataset. This new dataset is designed to support SR research on high-resolution thermal images in challenging conditions. A comparison of VGTSR2.0 with some common SR datasets is shown in Table~\ref{datacompare}.

\subsection{Data Collection and Analysis}
The VGTSR2.0 Dataset comprises 3,500 pairs of high-resolution optical and thermal images, acquired using DJI M30T and Matrice 3TD drones, the latter designed for the DJI Dock 2 platform. These drones are equipped with an uncooled VOx microbolometer sensor, enabling detailed thermal imaging with adjustable resolutions of 1080 $\times$ 1920 in standard mode and 1280 $\times$ 1024 in super-resolution mode.
To ensure alignment and authenticity of the imagery, all optical images are resized and cropped to match the 640 $\times$ 512 resolution of the thermal images. Manual alignment during the cropping process is performed to meet the requirements of multi-modal research methodologies, maintaining consistency across the dataset.

In the VGTSR2.0 dataset, we conduct a comprehensive statistical analysis of image pairs. As shown in Fig.~\ref{condition}, VGTSR2.0 includes a larger number of images captured under various lighting conditions, such as bright daylight, low-light environments at night, and complex light variations at dusk and dawn. This diversity effectively tests the adaptability and performance of OTUAV-SR algorithms in different lighting environments. Additionally, we capture images under varying concentrations of haze, from light to heavy, reflecting the challenges haze poses to image clarity and detail capture. The dataset also supports tasks such as low-light enhancement and dehazing. 
To enrich the dataset and enhance its challenge, the original VGTSR1.0 dataset is evenly split into three parts. Two parts are used to artificially generate foggy and low-light images, while the third part retains images captured under normal lighting conditions. We maintain a 2:1 ratio of real to synthetic data, ensuring the dataset's authenticity and reliability while expanding its diversity.
As shown in Fig.~\ref{sence}, the scenes in VGTSR2.0 are more diverse and no longer limited to campuses and streets. Specific scenes include huts, skyscrapers, car parks, farmland, ponds, lakes, streets, and schools, all captured at no less than 200 meters. This diversity and complexity present unique challenges for OTUAV-SR. The variety of architectural styles, vegetation, and topography, along with the significant differences in thermal properties between water bodies and land surfaces all contribute to complex thermal patterns, posing significant challenges for the OTUAV-SR task.
The data allocation for training the GDNet network on the VGTSR2.0 dataset is presented in Table\ref{train}. This allocation ensures a balanced dataset for evaluating super-resolution algorithms under diverse illumination and visibility conditions.
\begin{table*}
\renewcommand{\arraystretch}{1.8}

\caption{Comparison with state-of-the-art single image SR methods and guided image SR methods using both no-reference and reference evaluation metrics for $\times$4 and $\times$8 SR on the VGTSR2.0 dataset, employing bicubic (BI) and blur-downscale (BD) degradation models. Bold font indicates the best performance.}
\label{tab:vgtsr2.0-experiment}
\setlength{\tabcolsep}{3pt}	
{
\scriptsize 
\begin{tabular}{
c!{\vrule width 0.8pt}c|c|c|c!{\vrule width 0.8pt}c|c|c|c!{\vrule width 0.8pt}c|c|c|c!{\vrule width 0.8pt}c|c|c|ccccccccccccccccccc}
\toprule
\multirow{2}{*}{Method}  & \multicolumn{4}{c!{\vrule width 0.8pt}}{Scale:4 BI}           & \multicolumn{4}{c!{\vrule width 0.8pt}}{Scale:4 BD}            & \multicolumn{4}{c!{\vrule width 0.8pt}}{Scale:8 BI}           & \multicolumn{4}{c}{Scale:8 BD}
\\ \cline{2-17}                 
& PSNR$\uparrow$ & SSIM$\uparrow$ & LPIPS$\downarrow$ & NIQE$\downarrow$ & PSNR$\uparrow$ & SSIM$\uparrow$ & LPIPS$\downarrow$ & NIQE$\downarrow$
& PSNR$\uparrow$ & SSIM$\uparrow$ & LPIPS$\downarrow$ & NIQE$\downarrow$ & PSNR$\uparrow$ & SSIM$\uparrow$ & LPIPS$\downarrow$ & NIQE$\downarrow$   \\ 
\hline
Bicubic &26.76 &0.8054 &0.2000 &8.0723 &25.77 &0.7833 &0.2115 &8.2102 &22.74 &0.6364 &0.3736 &8.3001 &22.44 &0.6485 &0.3529 &7.8063 \\
EDSR\cite{lim2017enhanced}  &30.57 &0.8997 &0.0661 &6.0295 &30.38 &0.8978 &0.0658 &6.0750 &24.55 &0.7406 &0.2076 &7.5248 &24.54 &0.7462 &0.2080 &7.3880 \\
D-DBPN\cite{haris2018deep} &30.62 &0.9004 &0.0636 &5.9886 &30.44 &0.8976 &0.0638 &5.9680 &25.54 &0.7607 &0.1913 &7.2623 &24.62 &0.7523 &0.1968 &7.2018 \\
RCAN\cite{zhang2018image} &31.26 &0.9106 &0.0560 &6.1085 &31.06 &0.9080 &0.0556 &6.0534 &25.62 &0.7657 &0.1864 &7.2800 &24.75 &0.7544 &0.1978 &7.1194\\
RDN\cite{zhang2018residual} &31.10 &0.9080 &0.0549 &6.0649 &30.97 &0.9069 &0.0558 &5.9866 &25.58 &0.7615 &0.1783 &7.0791 &24.71 &0.7546 &0.1882 &7.0682 \\
SAN\cite{dai2019second}  &31.20 &0.9099 &0.0552 &6.0908 &31.09 &0.9084 &0.0530 &5.9830 &25.63 &0.7581 &0.1891 &7.0005 &24.76 &0.7649 &0.1814 &7.1269 \\
HAN\cite{niu2020single} &31.23 &0.9080 &0.0555 &6.0651 &30.81 &0.9041 &0.0580 &6.0333 &25.60 &0.7588 &0.1892 &7.1936 &24.78 &0.7595 &0.1971 &7.0373 \\
SwinIR\cite{liang2021swinir} &31.19 &0.9088 &0.0566 &6.0226 &31.06 &0.9089 &0.0571 &6.0376 &25.70 &0.7663 &0.1886 &7.2298 &24.86 &0.7618 &0.1923 &7.1302\\
Restormer\cite{zamir2022restormer} &30.99 &0.9053 &0.0577 &6.0493 &30.84 &0.9050 &0.0585 &6.0701 &25.73 &0.7677 &0.1842 &7.2032 &24.90 &0.7615 &0.1893 &7.0717\\
HAT\cite{chen2023activating}  &31.26 &0.9086 &0.0556 &6.0241 &31.10 &0.9079 &0.0554 &6.0125 &25.72 &0.7662 &0.1853 &7.2164 &24.89 &0.7616 &0.1881 &7.1401\\
UGSR\cite{gupta2021toward}&30.53 &0.8972 &0.0638 &5.9879 &30.34 &0.8962 &0.0651 &6.0119 &25.35 &0.7536 &0.1955 &7.5624 &24.64 &0.7483 &0.1948 &7.4004\\
MGNet\cite{zhao2023thermal} &31.33 &0.9116 &0.0531 &5.9544 &31.16 &0.9100 &0.0533 &5.9494 &26.12 &0.7829 &0.1634 &6.9192 &25.41 &0.7803 &0.1676 &6.9156 \\
CENet\cite{zhao2024modality}&31.42 &0.9146 &0.0517 &5.9764 &31.26 &0.9120 &0.0507 &5.9963 &26.28 &0.7877 &0.1573 &6.6822 &25.50 &0.7871 &0.1661 &6.7040\\ \hline
GDNet(Ours)&\textbf{31.51} &\textbf{0.9159} &\textbf{0.0495} &\textbf{5.9226} &\textbf{31.40} &\textbf{0.9134} &\textbf{0.0504 }&\textbf{5.8397} &\textbf{26.38 }&\textbf{0.7930 }&\textbf{0.1478} &\textbf{6.6799} &\textbf{25.67} &\textbf{0.7937} &\textbf{0.1562}&\textbf{6.6271} \\
\bottomrule
\end{tabular}

}
\end{table*}
\begin{table}
\centering
\renewcommand{\arraystretch}{1.5}
\caption{Quantitative comparison on real-world optics-guided dataset of VGTSR2.0 using no-reference evaluation metrics NIQE and BRISQUE, for ×4 SR employing the BI degradation model. Bold font indicates the best performance.}
\label{no reference}
\scriptsize 
\begin{tabular}{llcc|ccc}
\toprule
&Method        & NIQE $\downarrow$ & BRISQUE $\downarrow$ & Params(M) & FLOPs(G) \\ \hline
&Bicubic      & 8.2103            & 61.0891                       &-      & -        \\
&EDSR\cite{lim2017enhanced}         & 6.2160            & 49.2760                      &43M      &10.22G         \\
&D-DBPN\cite{haris2018deep}       & 6.1626            & 48.2979                       &10M     &44.28G         \\
&RCAN\cite{zhang2018image}         & 6.3056            & 48.0261                       &16M      &67.21G         \\
&RDN\cite{zhang2018residual}          & 6.2940            & 48.2174                       &22.3M      &52.36G         \\
&SAN\cite{dai2019second}          & 6.2929            & 47.6048                       &15.7M      &67.00G         \\
&HAN\cite{niu2020single}          & 6.2667            & 48.0174                       &15.4M      &123.85G         \\
&SwinIR\cite{liang2021swinir}       & 6.2161            & 47.7863                       &5M      &12.56G         \\
&Restormer\cite{zamir2022restormer}    & 6.2289            & 48.0228                       &25.3M      &23.79G         \\
&HAT\cite{chen2023activating}          & 6.2195            & 47.9008                       &5.2M      &13.18G         \\
&UGSR\cite{gupta2021toward}         & 6.1738            & 48.6573                       &4.5M      &46.60G         \\
&MGNet\cite{zhao2023thermal}        & 6.1328            & 47.1946                       &18.6M      &49.41G         \\
&CENet\cite{zhao2024modality}        & 6.1548            & 47.0924                       &11.8M      &28.89G         \\
&GDNet(Ours)  & \textbf{6.1018}    & \textbf{46.2755}               &11.9M     &32.07G         \\ \bottomrule
\end{tabular}
\end{table}
\subsection{Dataset Challenges}
The VGTSR2.0 dataset introduces several new challenges compared to the original dataset due to its unique characteristics:

\begin{itemize}
    \item \textbf{Image Quality Degradation:} Thermal images captured by UAV typically suffer from lower contrast and reduced texture details, while optical images are often affected by increased noise due to altitude and vibrations. These factors negatively impact the super-resolution process and potentially introduce artifacts.
    
    \item \textbf{Small Object Recovery:} UAV images often contain small objects, such as vehicles and pedestrians, making the recovery of edge details during SR tasks more challenging compared to simpler scenes captured by handheld cameras.
    
    \item \textbf{Environmental Variability:} The dataset comprises images captured under diverse lighting conditions (daylight, nighttime, and dawn/dusk) and varying haze concentrations (ranging from light to heavy), providing a comprehensive evaluation of the adaptability of SR algorithms.
    
    \item \textbf{Scene Diversity:} The dataset encompasses a wide range of scenes, including skyscrapers, parking lots, farmlands, ponds, lakes, streets, and schools, all captured at altitudes of at least 200 meters. This diversity in architectural styles, vegetation, and thermal properties of various surfaces adds to the complexity of SR tasks.
\end{itemize}

Furthermore, VGTSR2.0 incorporates both synthetic and real images, enhancing the dataset’s diversity and complexity. The synthetic images are produced through algorithms that simulate conditions such as fog and low light, offering a controlled yet demanding environment for testing algorithms.

\begin{figure*}
    \centering
    \includegraphics[width=.94\linewidth]{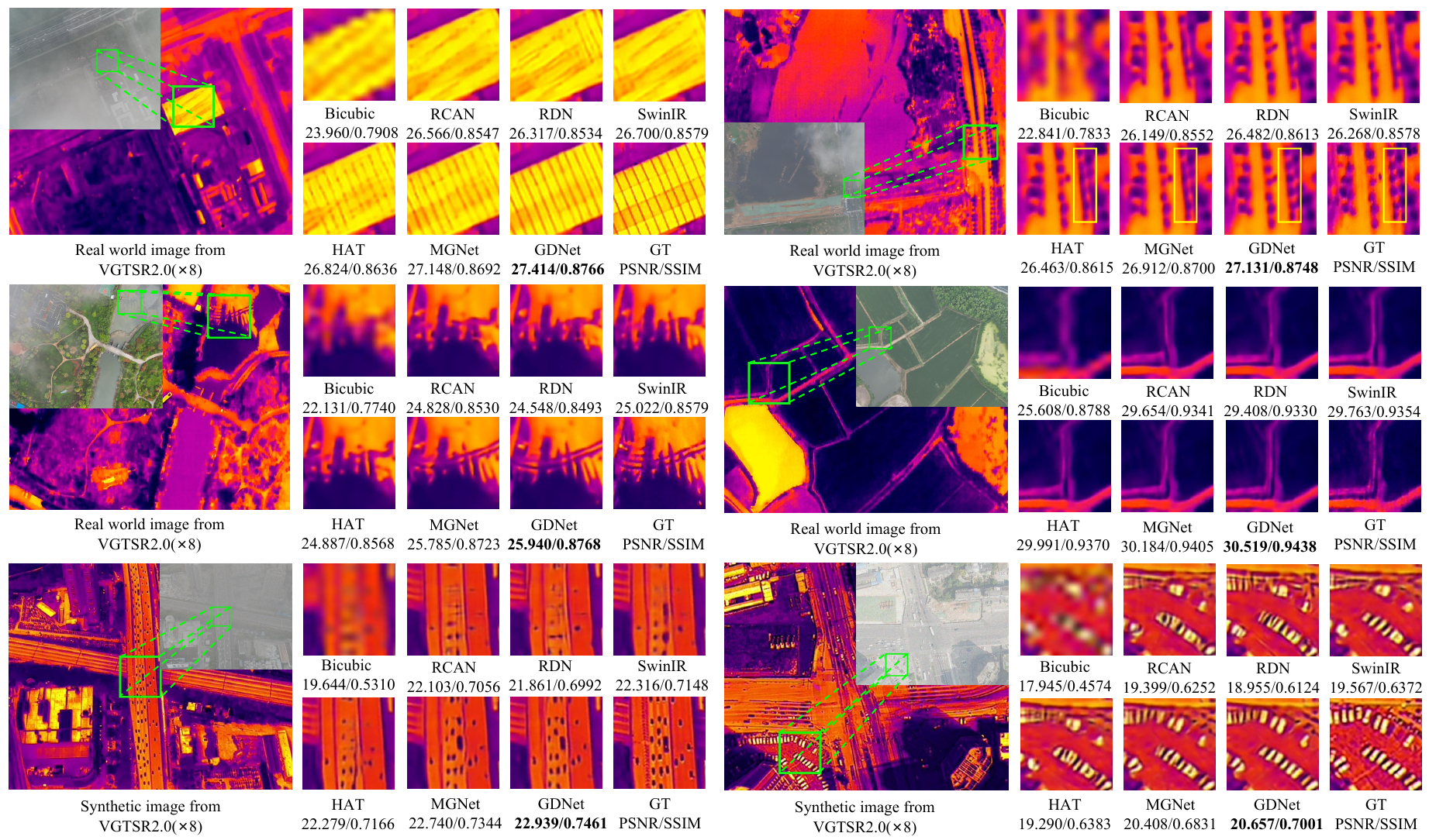}
    \caption{Qualitative comparison of super-resolution methods at $\times$8 SR in real-world and synthetic fog-obscured scenes. In foggy environments, the reconstruction results of GDNet exhibit clearer contours and demonstrate a visual performance that closely resembles the ground truth (GT) in both synthetic and real guided images.}
    \label{fog}
\end{figure*}
\begin{figure*}
    \centering
    \includegraphics[width=.88\linewidth]{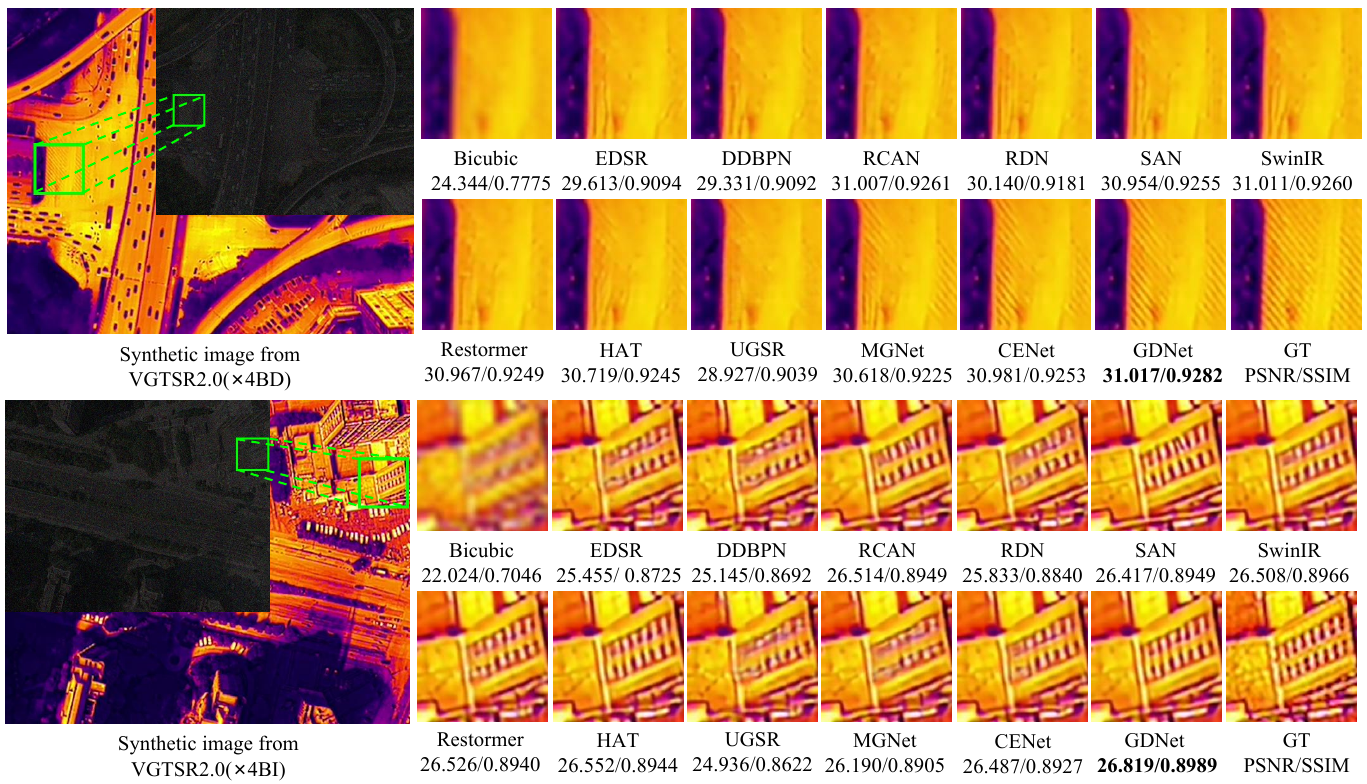}
    \caption{Qualitative comparison of super-resolution methods at $\times$4 SR in synthetic low-light scenes. Our proposed GDNet can still reconstruct better details with the guidance of optical images under these unfavorable conditions.}
  \label{4DarkSys}
\end{figure*}
\section{EXPERIMENTS}
In this section, the proposed GDNet and other state-of-the-art methods are comprehensively evaluated on the VGTSR2.0 dataset. First, the implementation details of GDNet will be presented in Section \ref{Implementation}. The effectiveness of our GDNet is then investigated in Section \ref{Quantitative}, and finally, ablation experiments and visualizations are reported in Section \ref{Ablations}.
\subsection{Implementation Details}
\label{Implementation}
We implement GDNet by PyTorch. For the optical image input branch under the three conditions, we use $640 \times 512$ optical images. The input for the thermal image input branch is low-resolution thermal images simulated by the degradation model. We train the entire network using the proposed three-stage training scheme. MOGM comprises four Residual Multiple Attention Groups (RMAGs), each containing one MGL, six STLs, one OMCL, and one OTL.
In our proposed GDNet, the kernel size, embed-dim, window size, attention head number and patch size are generally set to 3 $\times$ 3, 96, 8, 6, 1, respectively.
We set the resolution of the patches to be the same as in the SwinIR baseline and other super-resolution methods: 48 $\times$ 48.
For $\times$ 4 SR, a  48 $\times$ 48 pixel area is cropped from the LR thermal image, and a 192 $\times$ 192 pixel area from the corresponding position on the HR optical image is used as input. For $\times$ 8 SR, a 48 $\times$ 48 pixel area is similarly cropped from the LR thermal image, while a 384 $\times$ 384 pixel area is taken from the corresponding position on the HR optical image as input.
The model is trained with the Adam optimizer ($\beta_{1}$ = 0.9, $\beta_{2}$ = 0.99, and $\epsilon$ = 10$^{-8}$). The learning rate is initially set to 10$^{-4}$ and is then reduced by half every 200 epochs. The batch size is set to 8.

\begin{figure*}
    \centering
    \includegraphics[width=.85\linewidth]{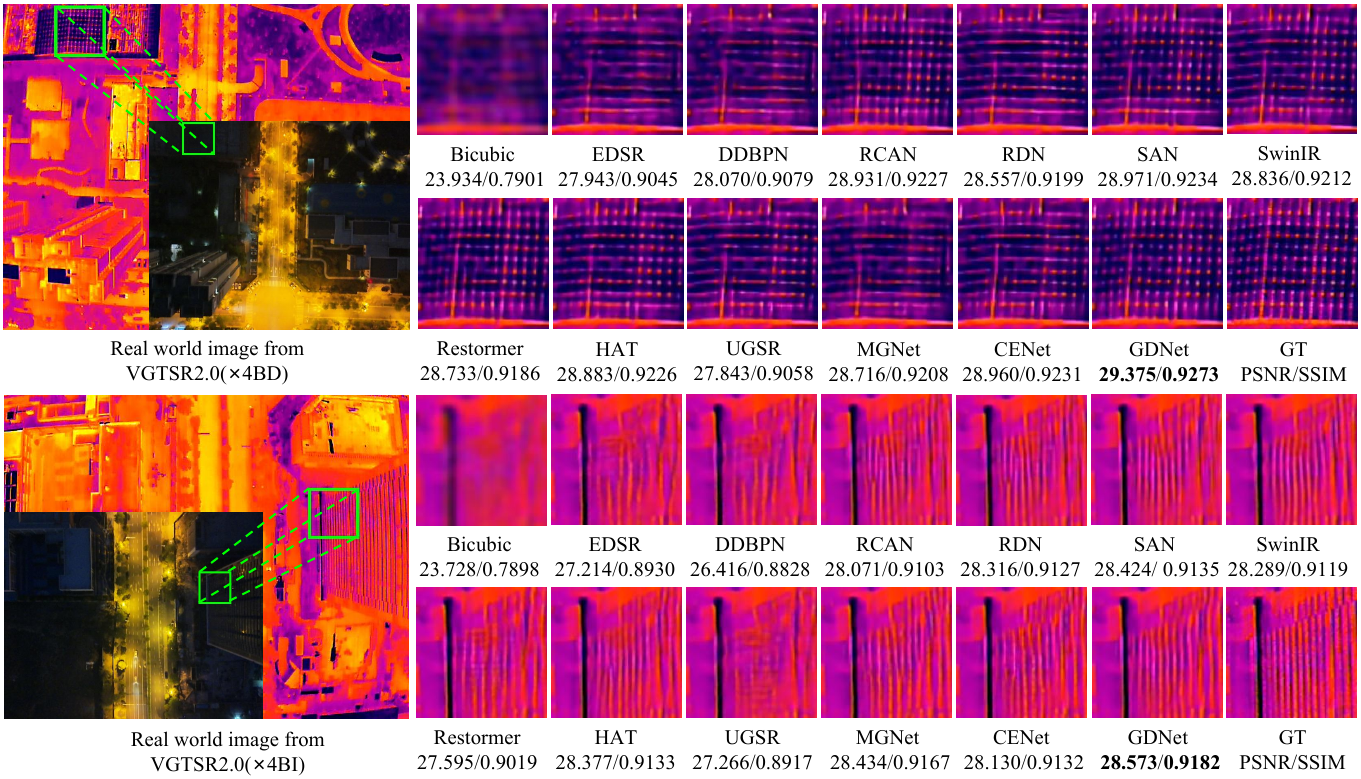}
    \caption{Qualitative comparison of super-resolution methods at $\times$4 SR in real-world low-light scenes. Under low-light conditions in real-world scenarios, the reconstruction results of the three guided methods USR, MGNet, and CENet are even inferior to those obtained from single image super-resolution methods. In contrast, our reconstruction results are closer to the ground truth.}
    \label{4realDark}
\end{figure*}
\begin{table*}
\renewcommand{\arraystretch}{1.8}

\caption{Comparison with state-of-the-art single image SR methods and guided image SR methods using reference evaluation metrics for $\times$4 and $\times$8 SR on a real-world optics-guided dataset of VGTSR2.0, employing bicubic (BI) and blur-downscale (BD) degradation models. Bold font indicates the best performance.}
\label{tab:Realword}
\setlength{\tabcolsep}{3pt}	
{
\scriptsize 
\begin{tabular}{
cc!{\vrule width 0.8pt}c|c|c!{\vrule width 0.8pt}c|c|c!{\vrule width 0.8pt}c|c|c!{\vrule width 0.8pt}c|c|cccccc}
\toprule
&\multirow{2}{*}{Method}  & \multicolumn{3}{c!{\vrule width 0.8pt}}{Scale:4 BI}           & \multicolumn{3}{c!{\vrule width 0.8pt}}{Scale:4 BD}            & \multicolumn{3}{c!{\vrule width 0.8pt}}{Scale:8 BI}           & \multicolumn{3}{c}{Scale:8 BD}
\\ \cline{3-15}                 
&& PSNR$\uparrow$ & SSIM$\uparrow$ & LPIPS$\downarrow$ & PSNR$\uparrow$ & SSIM$\uparrow$ & LPIPS$\downarrow$ 
& PSNR$\uparrow$ & SSIM$\uparrow$ & LPIPS$\downarrow$& PSNR$\uparrow$ & SSIM$\uparrow$ & LPIPS$\downarrow$&&  \\ 
\hline
&Bicubic &27.08 &0.8389 &0.2114 &26.08 &0.8232 &0.2224 &23.05 &0.7202 &0.3768 &22.77 &0.7265 &0.3525 \\
&EDSR\cite{lim2017enhanced}  &30.92 &0.9165 &0.0791 &30.71 &0.9145 &0.07899 &24.85 &0.7942 &0.2195 &24.84 &0.7936 &0.2197\\
&D-DBPN\cite{haris2018deep}  &31.01 &0.9178 &0.0765 &30.82 &0.9165 &0.0765 &25.97 &0.8096 &0.2007 &24.97 &0.7993 &0.2075 \\
&RCAN\cite{zhang2018image}&31.53 &0.9249 &0.0675 &31.31 &0.9229 &0.0671 &26.06 &0.8129 &0.1983 &25.08 &0.8026 &0.2094\\
&RDN\cite{zhang2018residual} &31.45 &0.9241 &0.0655 &31.33 &0.9234 &0.0664 &25.93 &0.8113 &0.1900 &24.92 &0.8019 &0.1998 \\
&SAN\cite{dai2019second} &31.51 &0.9248 &0.0663 &31.35 &0.9242 &0.0641 &25.01 &0.8046 &0.2001 &26.02 &0.8128 &0.1925\\
&HAN\cite{niu2020single}&31.54 &0.9247 &0.0669 &31.20 &0.9214 &0.0694 &26.05 &0.8113 &0.2010 &25.11 &0.8026 &0.2086 \\
&SwinIR\cite{liang2021swinir} &31.50 &0.9245 &0.0680 &31.36 &0.9239 &0.0688 &26.14 &0.8139 &0.2000 &25.22 &0.8070 &0.2046\\
&Restormer\cite{zamir2022restormer} &31.29 &0.9223 &0.0696 &31.12 &0.9208 &0.0705 &26.14 &0.8155 &0.1951 &25.20 &0.8082 &0.2003\\
&HAT\cite{chen2023activating}  &31.54 &0.9252 &0.0666 &31.40 &0.9241 &0.0665 &26.13 &0.8147 &0.1967 &25.17 &0.8082 &0.1989\\
&UGSR\cite{gupta2021toward}&30.90 &0.9169 &0.0761 &30.63 &0.9144 &0.0775 &25.73 &0.8035 &0.2056 &24.87 &0.7991 &0.2061\\
&MGNet\cite{zhao2023thermal} &31.62 &0.9286 &0.0638 &31.46 &0.9275 &0.0640 &26.53 &0.8330 &0.1741 &25.75 &0.8299 &0.1790\\
&CENet\cite{zhao2024modality}&31.73 &0.9308 &0.0624 &31.57 &0.9304 &0.0604 &26.66 &0.8390 &0.1686 &25.80 &0.8316 &0.1776\\ \hline
&GDNet(Ours)&\textbf{31.80} &\textbf{0.9325} &\textbf{0.0596} &\textbf{31.71} &\textbf{0.9320 }&\textbf{0.0588} &\textbf{26.83 }&\textbf{0.8447} &\textbf{0.1579} &\textbf{26.01} &\textbf{0.8392} &\textbf{0.1671} \\
\bottomrule
\end{tabular}
}
\end{table*}
\begin{figure*}
    \centering
    \includegraphics[width=0.80\linewidth]{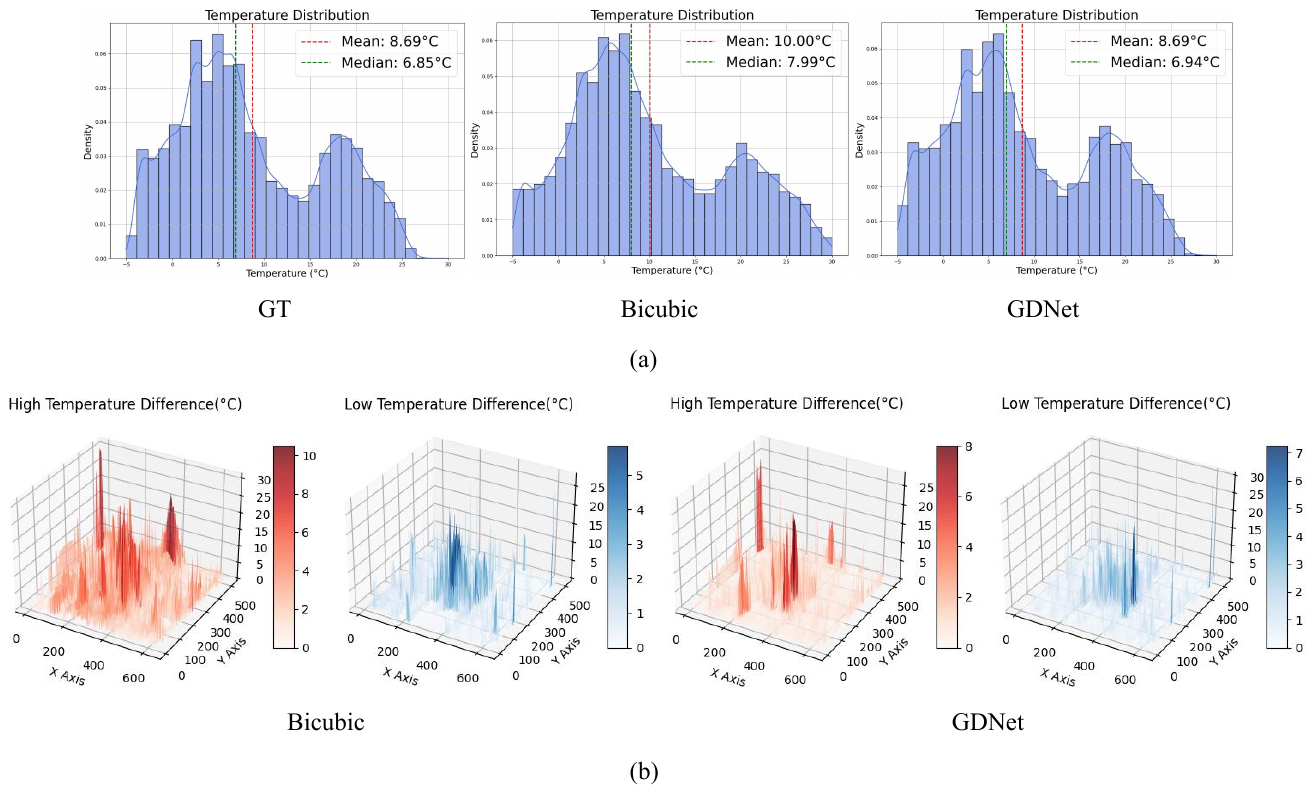}
    \caption{Visual comparison of temperature distribution in SR results from different methods: (a) Histogram of temperature distribution, with the red and green dashed lines representing the mean and median temperatures of the reconstructed results, respectively. (b) 3D plot of temperature difference, showing the difference between the temperature of the recovered image and that of the GT.}
    \label{temputure}
\end{figure*}

\begin{figure*}
    \centering
    \includegraphics[width=.85\linewidth]{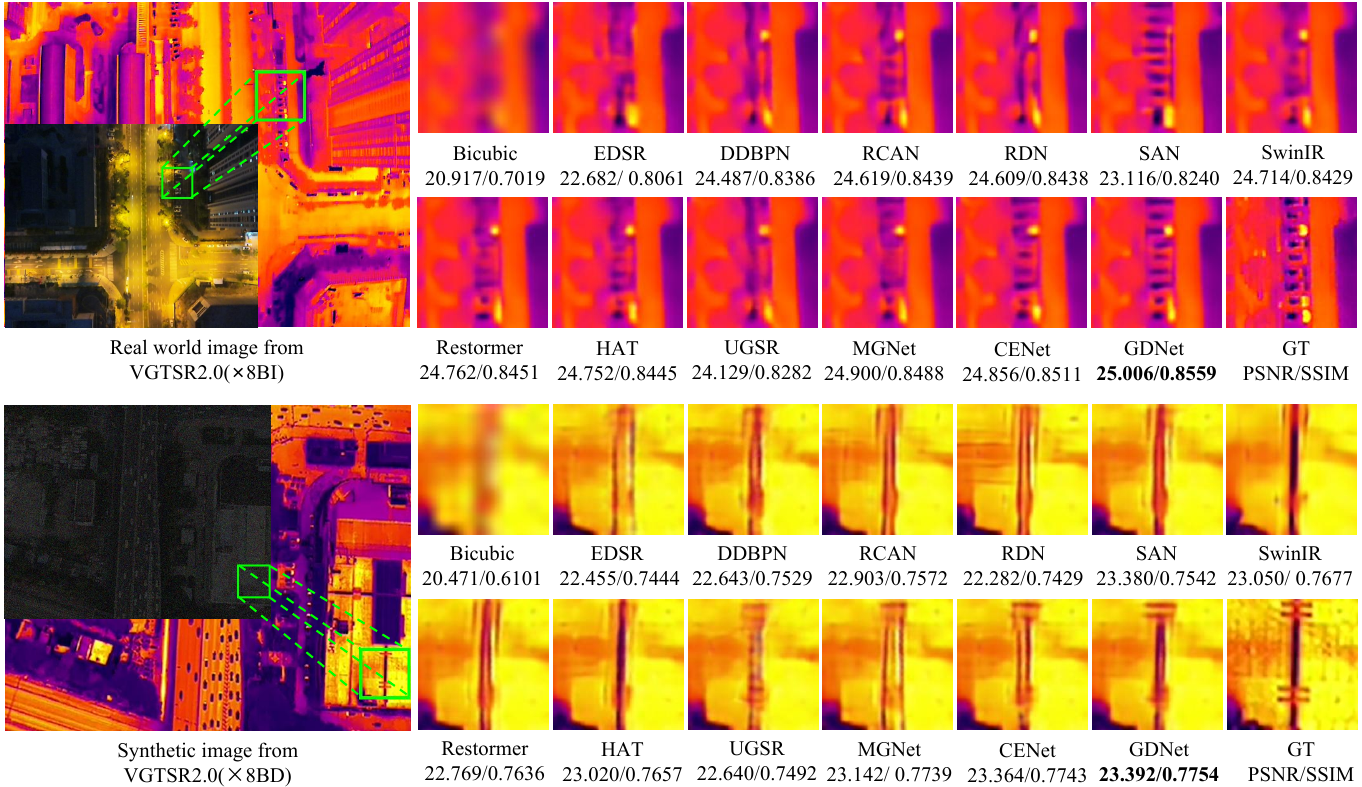}
    \caption{Qualitative comparison of super-resolution methods at $\times$8 SR in both real-world and synthetic low-light scenes. Our method also achieves better results on the 8$\times$8 degradation model.}
    \label{8BIBDdark}
\end{figure*}
\subsection{Quantitative and Qualitative Evaluation}
\label{Quantitative}
To validate the effectiveness of GDNet, we performed $\times$4 and $\times$8 super-resolution experiments on the VGTSR2.0 dataset. We report the results with eight state-of-the-art SISR methods, including CNN-based models D-DBPN~\cite{haris2018deep}, RDN~\cite{zhang2018residual}, SAN~\cite{dai2019second}, EDSR~\cite{lim2017enhanced}, RCAN~\cite{zhang2018image}, HAN~\cite{niu2020single}; and Transformer-based models SwinIR~\cite{liang2021swinir}, HAT~\cite{chen2023activating}, and Restormer~\cite{zamir2022restormer}. Additionally, we compare three OTUAV-SR methods: UGSR~\cite{gupta2021toward}, MGNet~\cite{zhao2023thermal}, and CENet~\cite{zhao2024modality}, with CENet currently achieving the best performance.
We saved the best image results for all models and consistently assessed image quality using both Reference and No-Reference evaluation metrics.

Table~\ref{tab:vgtsr2.0-experiment} presents the results for $\times$4 and $\times$8 SR with BD and BI degradation models on the whole VGTSR2.0 dataset.
For $\times$4 SR, our method achieves a 0.09 dB gain in PSNR, a 0.0013 increase in SSIM with the BI model, and a 0.14 dB and 0.0014 improvement with the BD model. For $\times$8 SR, the BI model yields a 0.1 dB PSNR and 0.0053 SSIM enhancement, while the BD model sees a 0.17 dB PSNR and a 0.0066 SSIM increase. Additionally, our method achieves the best performance in both LPIPS and NIQE metrics, demonstrating its incremental advantages across different scales and metrics.
Table~\ref{tab:Realword} presents the quantitative results for PSNR, SSIM, and LPIPS, with bold text indicating the best performance for each metric. GDNet achieves the highest average PSNR, SSIM, and the lowest LPIPS across both degradation types at $\times$4 and $\times$8 SR. Specifically, for $\times$4 SR, GDNet achieves PSNR gains of 0.07 dB and 0.14 dB in the BI and BD models, respectively. For $\times$8 SR, the PSNR increased by 0.17 dB in the BI degradation model and by 0.21 dB in the BD degradation model, underscoring the effective performance of GDNet in real optical-guided thermal SR.

Different single image super-resolution methods exhibit closely comparable performance, however, transformer-based SR methods generally outperform those based on CNN. In guided super-resolution, UGSR introduces noise by transferring textures from optical images through fusion methods, leading to lower PSNR and SSIM. MGNet and CENet respectively incorporate edge, semantic, and appearance cues, alongside modality conversion and task-assisted super-resolution, offering significant advantages over other methods. By decoupling the optical representation and utilizing adaptive fusion, our GDNet achieves optimal performance.
\begin{figure*}
    \centering
    \includegraphics[width=0.76\linewidth]{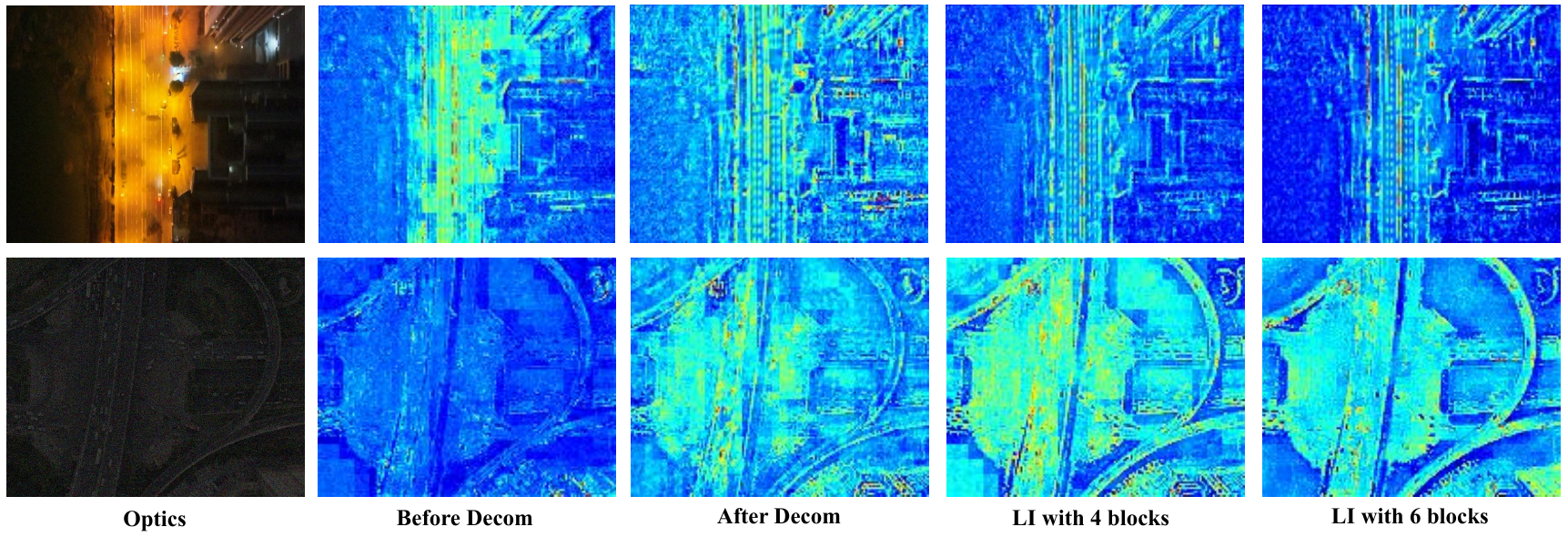}
    \caption{Optical feature visualization in low-light scenes before and after using the Decom and LI branches. The first row displays the feature visualization results of real optical images, while the second row shows the feature visualization results of synthetic optical images.}
    \label{darkview}
\end{figure*}
\begin{figure*}
    \centering
    \includegraphics[width=0.98\linewidth]{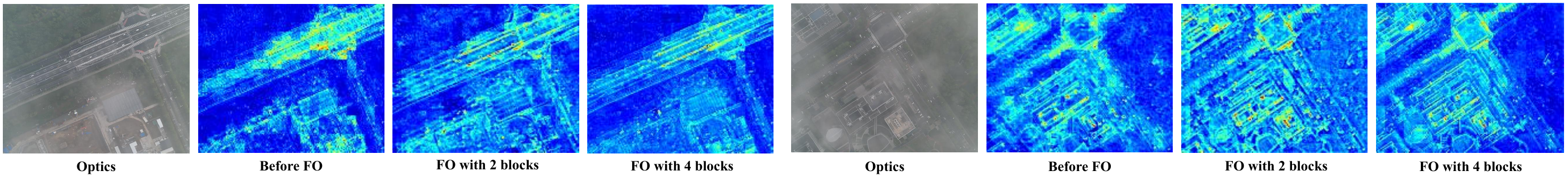}
    \caption{Optical feature visualization in fog-obscured sceneS before and after using the FO branch. After processing the optical features under fog conditions in the FO module, the feature structure becomes clearer, and the noise is significantly reduced.}
    \label{fogview}
\end{figure*}
To further validate the effectiveness and adaptability of our method, we perform SR on thermal images guided by real optical images captured from VGTSR2.0. We conduct no-reference image quality assessments using NIQE \cite{mittal2012making} and BRISQUE \cite{mittal2012no}, with the quantitative results summarized in Table~\ref{no reference}. The experimental results demonstrate that our GDNet significantly improves performance in both NIQE and BRISQUE metrics compared to other methods. Compared to other single image super-resolution models and guided super-resolution models, our model demonstrates advantages in terms of parameter efficiency. Table~\ref{no reference} also presents a comparison of different SR methods based on model size, FLOPs, and performance. Compared to MGNet, our model reduces the parameter count by 36\%. These results indicate that our approach not only maintains a comparable parameter scale to the leading model, CENet, but also achieves superior performance. 

Our model demonstrates superior performance not only on evaluation metrics such as PSNR and SSIM but also exhibits significant improvements in perceptual quality.
As illustrated in Fig.~\ref{fog}, single image SR methods struggle to recover detailed textures, and guided image SR methods face challenges when dealing with unknown textures and appearance. In contrast, our proposed GDNet can still reconstruct better details using optical images under foggy conditions. As shown in Fig.~\ref{4DarkSys}, ~\ref{4realDark}, and ~\ref{8BIBDdark}, our proposed GDNet recovers textures that are closest to the ground truth thermal images, even under low-light conditions for both BD and BI degradation models, while optics-guided thermal super-resolution methods even introduce erroneous texture information. Particularly noteworthy is the substantial enhancement achieved by our GDNet in $\times$8 super-resolution of drone thermal images degraded by BD models. 
Furthermore, the SR results obtained by GDNet exhibit the closest temperature distribution to the GT. As shown in Fig.~\ref{temputure}(a), GDNet achieves high-quality results with an average temperature value identical to the GT and a median difference of less than 1°C. Additionally, compared to the Bicubic model, GDNet achieves a smaller temperature difference from the GT, as shown in Fig.~\ref{temputure}(b).


\begin{table}
\renewcommand{\arraystretch}{1.5}
\centering
\scriptsize

\caption{Ablation study of the AGM in processing the corresponding attributes of VGTSR2.0. The best results obtained using a single branch for each attribute are highlighted in bold above the horizontal line in the table, and the results obtained using multiple branches that outperform a single branch are highlighted in bold below the horizontal line.}
\resizebox{\linewidth}{!}{%
\begin{tabular}{c| c c c }
\toprule
Method        &Normal &Fog &Low-light\\ \hline
Baseline      &28.64/0.8895    &33.69/0.9399   &30.66/ 0.9355    \\
MGNet      &29.24/0.8985       &33.70/0.9400   &30.56/0.9324  \\
NC        &\textbf{29.26/0.9065 }      &33.51/0.9376   &29.81/0.9274   \\
FO        &28.22/0.8896      &\textbf{33.74/0.9409}  &30.03/0.9303  \\
LI        &26.97/0.8604      &33.11/0.9268 & \textbf{30.68/0.9353} \\ \hline
NC+LI  &29.25/0.8988      &33.64/0.9352  &\textbf{30.70/0.9354} \\
NC+FO  &\textbf{29.27}/0.9023      &\textbf{33.74/0.9411}  &30.13/0.9312  \\
LI+FO  & 28.40/0.8881     &\textbf{33.75}/0.9403  &\textbf{30.76/0.9358}  \\
NC+FO+LI   & 29.24/\textbf{0.9073}     & \textbf{33.78/0.9418} & \textbf{30.83/0.9365} \\ \bottomrule
\end{tabular}}
\label{ablationofagm}
\end{table}
\subsection{Ablation Experiments}
\label{Ablations}

\begin{figure*}
    \centering
    \includegraphics[width=.82\linewidth]{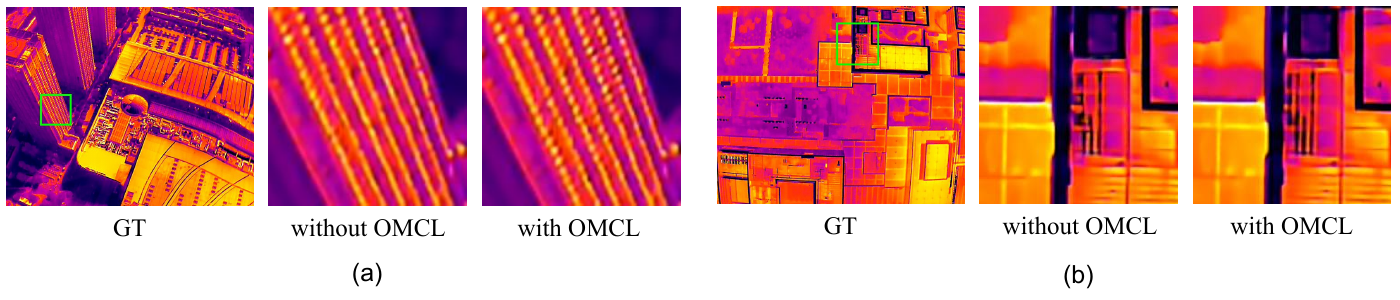}
    \caption{Visual comparison of × 8 SR results with and without OMCL on the VGTSR2.0 normal test set. (a) Thermal image of a skyscraper scene. (b) Thermal image of a bungalow scene.}
    \label{omclvisual}
\end{figure*}

\begin{figure*}
    \centering
    \includegraphics[width=.82\linewidth]{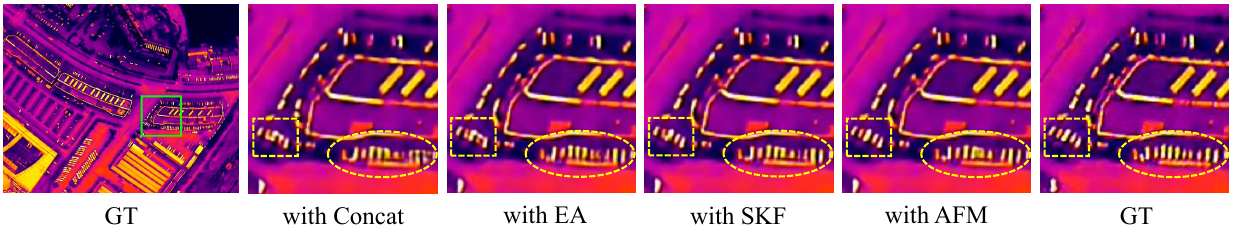}
    \caption{Visual comparison of $\times$8 SR results using different fusion modules on the VGTSR2.0 test set. AFM generates superior visual outcomes.}
    \label{ronghe}
\end{figure*}

\noindent\textbf{Effectiveness of Attribute-specific Guidance Module.}
To further quantitatively assess the efficacy of each attribute-specific branch in handling the corresponding scene attributes, we conduct a comparative study using attribute-based VGTSR2.0. As shown in Table~\ref{ablationofagm}, the terms Normal, Fog, and Low-light represent the datasets for normal light, foggy, and low light conditions, respectively. Each attribute-specific branch achieves superior performance on its corresponding attribute, indicating that our GDNet can effectively construct optical appearance representations tailored to different weather conditions by leveraging attribute information. Furthermore, we observe that the combined results from multiple branches are at least comparable to those obtained from individual branches. This suggests that designing branches for different attributes effectively addresses their respective optical appearance challenges under varying weather conditions, while the integration of multiple branches effectively mitigates optical representation challenges across diverse weather conditions.

We simultaneously visualize the intermediate features of optical information processed through the LI and FO branches. As shown in Fig.~\ref{darkview}, influenced by the street light's exposure, the road's detailed information is obscured, and Decom produces a clearer structure, while LI further reduces the noise. As illustrated in Fig.~\ref{fogview}, optical features in foggy conditions are challenging to render before using FO. However, incorporating 4-layer blocks yields optical features with reduced noise and enhanced texture sharpness, demonstrating the effectiveness of the FO module.

\begin{table}
\renewcommand{\arraystretch}{1.5}
\scriptsize
\begin{center}
\caption{Ablation Study of the STL, MGL, OMCL, and OCL layers in the MOGM Component}
\begin{tabular}{cccccccccccc} 
\toprule
STL &MGL & OMCL &OTL &Scale& PSNR$\uparrow$  & SSIM$\uparrow$  &NIQE$\downarrow$ \\ \hline
       $\checkmark$& && & 4     &31.06  &0.9089&6.0376  \\
      $\checkmark$ &$\checkmark$ && & 4     &31.31 &0.9127&5.8559 \\
 $\checkmark$ &$\checkmark$ &$\checkmark$ && 4     &31.38& \textbf{0.9135}&5.8517\\
  $\checkmark$ &$\checkmark$ &$\checkmark$ &$\checkmark$  & 4     & \textbf{31.40} & 0.9134&\textbf{5.8397} \\ \bottomrule
\end{tabular}
\label{OMCL}
\end{center}
\end{table}
\begin{table}[!ht]
\renewcommand{\arraystretch}{1.35}
\scriptsize
\caption{Ablation study of AFM on the VGTSR2.0 dataset. Concat, EA, and SKF denote Concatenation, Element-Wise Addition and SKFusion, respectively.(SR Scale Factor 4).}
\begin{center}
\begin{tabular}{cc|ccccc}
\toprule
       &Fusion Method &         & PSNR$\uparrow$&SSIM$\uparrow$  &NIQE$\downarrow$& \\ \hline
       &Concat&           & 31.21    &0.9080&5.9992\\
        &EA &            & 31.35  &0.9123  & 5.9374 \\
       &SKF &            &31.37  &0.9131&5.8621 \\
       &APM         &  &         \textbf{31.40}   & \textbf{0.9134} &\textbf{5.8397}\\
\bottomrule
\end{tabular}
\label{APM}
\end{center}
\end{table}
\noindent\textbf{Effectiveness of Fusion Module and OMCL.}
To validate the effectiveness of the proposed AFM, we compare various fusion methods, including element-wise addition, concatenation, and an improved SKFusion based on SKNet for integrating guidance features from different branches. The results, presented in Table~\ref{APM}, indicate that our AFM achieves the best performance. As illustrated in Fig.~\ref{ronghe}, the vehicle profile recovered using the AFM fusion module is noticeably clearer. To further evaluate the effectiveness of OMCL and compare it with the original OTL, we perform addition and deletion operations on OMCL. The experimental setup is outlined in Table~\ref{OMCL}. Compared to MGL+STL, the addition of OMCL yields improvements of 0.07 in PSNR and 0.0008 in SSIM. Additionally, incorporating OTL at the end, following the HAT method, resulted in only a 0.02 increase in PSNR. Our designed OMCL delivers significantly better performance. In visualization, the SR results with OMCL contain more detail and texture information, as shown in Fig.~\ref{omclvisual}.
\begin{figure*}
    \centering
    \includegraphics[width=.85\linewidth]{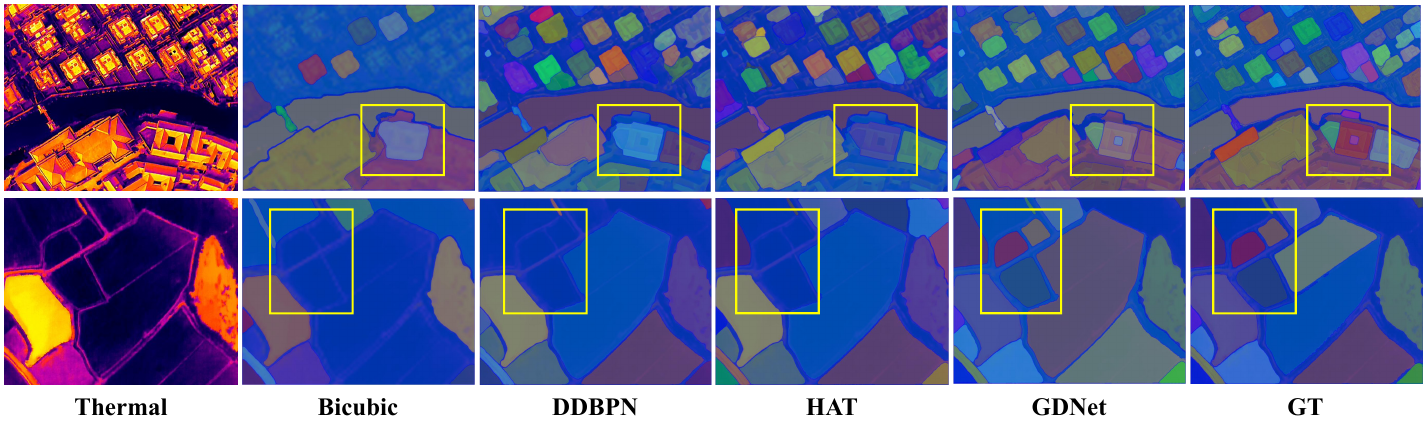}
    \caption{Visual comparison of segmentation results after SR using different methods. The first row represents the segmentation of building images, while the second row illustrates the segmentation of parcel images, both based on the SAM~\cite{kirillov2023segment}.}
    \label{sam}
\end{figure*}
\begin{figure*}
    \centering
    \includegraphics[width=.85\linewidth]{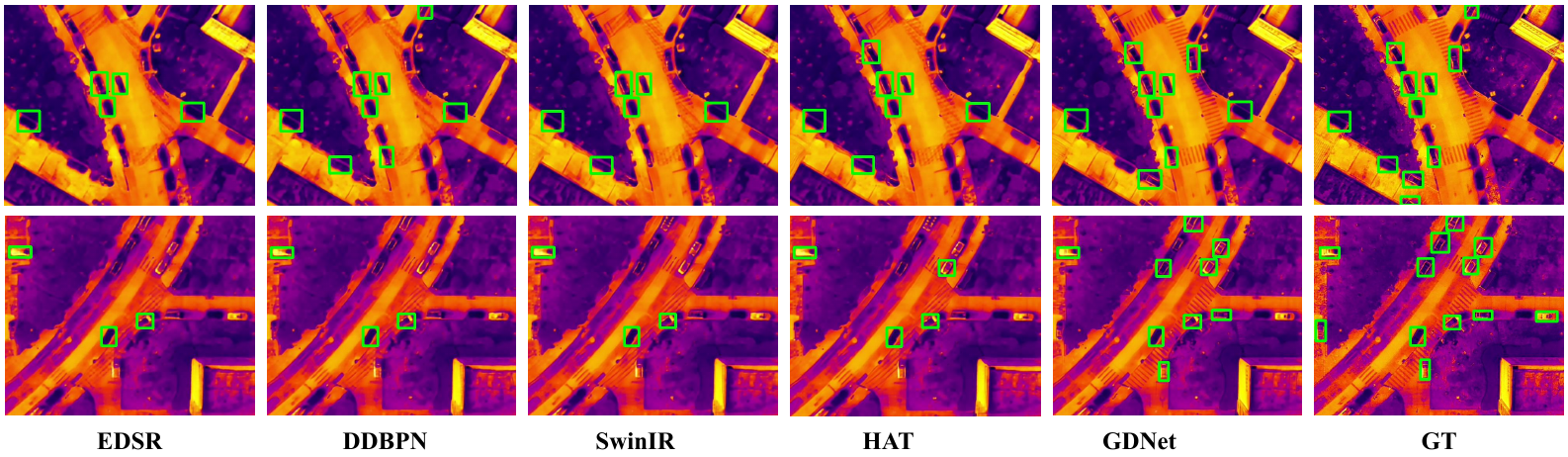}
    \caption{Visual comparison of object detection results after SR using different methods. Comparison indicates that the reconstruction results from GDNet outperform those of other methods in downstream detection tasks.}
    \label{detect}
\end{figure*}
\subsection{Comparison on Downstream Tasks}
To validate the practicality of GDNet, we conduct a series of experiments on downstream tasks, specifically focusing on semantic segmentation and object detection on SR results. As illustrated in Fig.~\ref{sam}, our method yields results that are close to ground truth in urban building segmentation and cultivated land parcel extraction. Furthermore, as shown in Fig.~\ref{detect}, GDNet successfully detects a greater number of target vehicles compared to current mainstream methods, such as HAT and SwinIR, which exhibit a higher rate of missed detections. These findings demonstrate GDNet's superiority in downstream tasks, including semantic segmentation and object detection.

\section{Conclusion}
In this paper, we introduce VGTSR2.0, a comprehensive benchmark dataset for optics-guided thermal UAV image super-resolution (OTUAV-SR). This dataset comprises 3,500 well-aligned optical-thermal image pairs captured under diverse, challenging conditions and scenes. To address the limitations of existing methods in generating effective guidance features across various UAV conditions, we propose a novel Guidance Disentanglement Network (GDNet). The GDNet incorporates an Attribute-specific Guidance Module (AGM) with specialized branches for low-light, foggy, and normal lighting conditions, ensuring robust performance across diverse environments. Furthermore, our Attribute-aware Fusion Module (AFM) adaptively integrates optical features in complex scenes by extracting information from three disentangled feature spaces. Quantitative and qualitative experiments demonstrate that GDNet significantly outperforms state-of-the-art single image super-resolution and OTUAV-SR methods on the VGTSR2.0 dataset, while also exhibiting strong performance in downstream tasks. In future study, we will further explore non-aligned guided super-resolution method at different resolutions.

\section*{Declaration of Competing Interest}
The authors declare that they have no known competing financial interests or personal relationships that could have appeared to influence the work reported in this paper.

\section*{Acknowledgement}
This work was supported in part by the National Natural Science Foundation of China(No. 62306005, 62006002, and 62076003), in part by the Joint Funds of the National Natural Science Foundation of China(No. U20B2068), in part by the Natural Science Foundation of Anhui Province(No. 2208085J18 and 2208085QF192), and in part by the Natural Science Foundation of Anhui Higher Education Institution (No. 2022AH040014).

\bibliographystyle{unsrt}

\bibliography{ref}

\end{sloppypar}
\end{document}